\documentclass[12pt]{article}
 
\usepackage[curve,matrix,arrow]{xy}

\usepackage{amsbsy,amssymb,amsmath,bm}
\usepackage{color,epsfig,rotate}

\usepackage{graphicx}

\usepackage{subfigure}

\usepackage{epsf}
\usepackage{epsfig}
\usepackage{amssymb}
\usepackage{amsthm}
\usepackage{amsmath}
\usepackage{epstopdf}

\usepackage{pstricks,pst-fill,pst-text,pst-node,pstricks-add}
\usepackage{psfrag}

\def\be{\begin{equation}}
\def\ee{\end{equation}}
\def\ba{\begin{eqnarray}}
\def\ea{\end{eqnarray}}

\newcommand{\lra}{{\ \longrightarrow\ }}

\input{xy}
\xyoption{all}

\newcommand{\rme}{{\rm e}}

\newcommand{\stf}{t}

\newcommand{\thl}{\psset{xunit=2mm,yunit=2mm}\begin{pspicture}(0,0)(0.75,1)
\psline[linecolor=black,linewidth=1.0pt](0,-0.5)(0,1)
\end{pspicture}}

\newcommand{\fug}{n}

\newcommand{\one}{\boldsymbol{1}}
\newcommand{\tensor}{\otimes}

\newcommand{\fus}{\times_{f}}
\newcommand{\q}{\mathfrak{q}}
\newcommand{\ffrac}[2]{\mbox{\footnotesize$\displaystyle\frac{#1}{#2}$}}

\newcommand{\TLq}[1]{TL_{\q,#1}}

\newcommand{\inv}{\mathrm{inv}}
\newcommand{\modd}{\,\mathrm{mod}\,}

\newcommand{\dd}{\mathsf{d}}

\newcommand{\cent}{\mathfrak{Z}}

\newcommand{\LQG}{U_{\q} s\ell(2)}

 \newcommand{\leftact}{\triangleright}
  \newcommand{\rightact}{\triangleleft}

\newcommand{\Hilb}{\mathcal{H}}

\newcommand{\veven}[1]{|v^{\text{even}}\rangle}
\newcommand{\vodd}[1]{|v^{\text{odd}}\rangle}

\newcommand{\vacr}{|\text{vac}\rangle}

\newcommand{\oN}{\mathbb{N}}

\newcommand{\Vir}{\mathfrak{vir}}

\newcommand{\gl}{g\ell}

\newcommand{\modWeyl}{\mathcal{W}}
\newcommand{\modWeylj}[1]{\modWeyl_{#1}}

\newcommand{\myar}{\ar@{-->}@[|(1.0)]}

\newcommand{\IrrTL}[1]{\mathsf{X}_{#1}}
\newcommand{\tIrrTL}[1]{\tilde{\mathsf{X}}_{#1}}
\newcommand{\PrTL}[1]{\mathsf{T}_{#1}}
\newcommand{\StTL}[1]{\mathsf{S}_{#1}}
\newcommand{\StTLn}[2]{\mathsf{S}_{#1}[#2]}
\newcommand{\tStTL}[1]{\tilde{\mathsf{S}}_{#1}}

\newcommand{\Wlat}{\mathcal{W}}
\newcommand{\Wlatq}[2]{\Wlat_{#1,#2}}
\newcommand{\modwX}{\mathcal{X}}

\newcommand{\modwY}{\mathcal{Y}}

\newcommand{\VK}{\mathcal{K}}
\newcommand{\VP}{\mathcal{T}}

\newcommand{\Verma}{\mathcal{V}}

\newcommand{\Ket}[1]{\left|#1  \right>}

\newcommand{\Braket}[1]{\left<#1  \right>}

\renewcommand{\geq}{\,{\geqslant}\,}
\renewcommand{\leq}{\,{\leqslant}\,}
\renewcommand{\ge}{\,{\geqslant}\,}
\renewcommand{\le}{\,{\leqslant}\,}

\theoremstyle{definition}

\newcommand{\BX}{\mathcal{X}}

\newcommand{\BW}{\mathcal{W}}

  \newcommand{\ssp}{\mathfrak{sp}}
  \newcommand{\spinf}{\ssp_{\infty}}
\newcommand{\interchalg}{\mathcal{S}}
\newcommand{\enrg}{S}
\newcommand{\bz}{\bar{z}}

\newcommand{\der}{\partial}
\newcommand{\bder}{\bar{\partial}}

\begin{document}

\title{Logarithmic Conformal Field Theory: a Lattice Approach}


\author{\small  A.M. Gainutdinov$^{1}$, J.L. Jacobsen$^{2,3}$, N. Read$^{4}$, H. Saleur$^{1,5}$ and R. Vasseur$^{1,2}$\\
 \small ${}^1$Institut de Physique Th\'eorique, CEA Saclay,
  91191 Gif Sur Yvette, France \\
 \small ${}^2$LPTENS, 24 rue Lhomond, 75231 Paris, France \\
 \small ${}^3$ Universit\'e Pierre et Marie Curie, 4 place Jussieu, 75252 Paris, France \\
 \small ${}^4$ Department of Physics, Yale University, P.O. Box 208120,
New Haven, CT 06520-8120, USA \\
 \small ${}^5$ Department of Physics, University of Southern California, Los Angeles, CA 90089-0484, USA}

\maketitle


\begin{abstract}

Logarithmic Conformal Field Theories (LCFT) play a key role, for instance,  in the description of  critical geometrical problems (percolation, self avoiding walks, {\it etc.}), or of 
critical points in several classes of disordered systems (transition between plateaus in the integer and spin quantum Hall effects).
Much progress in their understanding  has been obtained by studying algebraic features of their lattice regularizations. For reasons which are not entirely understood, the non semi-simple associative algebras underlying these lattice models -- such as the Temperley--Lieb algebra or the blob algebra -- indeed exhibit, in finite size, properties that are in full correspondence with those of their continuum limits. This applies to the structure of indecomposable modules, but also to fusion rules, and provides an `experimental' way of measuring couplings, such as the `number $b$' quantifying the logarithmic coupling of the stress energy tensor with its partner. Most results obtained so far have concerned boundary LCFTs, and the associated indecomposability in the chiral sector. While the bulk case is considerably more involved (mixing in general left and right moving sectors), progress has also been made  in this direction recently, uncovering fascinating structures. This article provides a short general review of our work in this area.

\end{abstract}

\section{Introduction}


While the tools and ideas of conformal field theory (CFT) \cite{BPZ} have become standard in low dimensional condensed matter physics, few of the fully solved, minimal unitary CFTs have actually found realistic applications. Out of the famous series with central charges given by~\cite{FQS,Cardy}
\begin{equation}
c=1-{6\over m(m+1)} ,~\quad m\hbox{ integer}\;\geq 3,\label{cunit}
\end{equation}
for instance, only the very few first values of $m$ correspond to experimentally (or numerically) observable critical points in statistical mechanics. This is because, as $m$ increases, more and more relevant operators are allowed \cite{Zamo} which are not constrained by symmetries, requiring the un-realistic fine tuning of more and more parameters. The situation has been somewhat more favorable in the field of quantum impurity problems (related to boundary conformal field theory~\cite{Cardybdr}), where for instance a lot of results for the  $SU(2)_k$ WZW theory have found applications in our understanding of the $k$-channel Kondo problem~\cite{AffLud}. Yet, observing experimentally even the two-channel Kondo problem remains difficult~\cite{Delft}.

Most physical applications of conformal invariance have involved instead CFTs whose understanding is not complete. This includes two dimensional geometrical problems like  self-avoiding walks and percolation, where the statistical properties of large, scaling objects are known \cite{Saleur87} to be described by critical exponents, correlation functions, {\it etc.}, pertaining to CFTs with central charge $c=0$. Such theories are necessarily {\sl not unitary} , since the only unitary CFT with $c=0$ admits, as its unique observable, the identity field, with conformal weight $h=\bar{h}=0$~\cite{Cardy}. 

Non-unitarity is certainly unpleasant from a field theoretic point of view: it  corresponds roughly (for more precise statements see below) to dealing with `hamiltonians' that are not Hermitian, and is probably non-sensical in the context of particle physics applications. In statistical mechanics, however, non-unitarity is rather common. In the case of polymers or percolation, it occurs because the basic problems one is interested in are non-local in nature -- in percolation for instance, an important observable (the order parameter) is related with the probability that a cluster connects two points far apart, while wandering without limits  through the system. 

This non-locality is easily traded for a local formulation which, however, involves complex Boltzmann weights~\cite{Nienhuis}. In the theory of self-avoiding walks for instance, one wants to cancel loops, which can be done by allowing elementary steps on the edges of a honeycomb lattice, and giving to each left/right turn a {\sl complex weight} $e^{\pm i\pi/12}$. Since for a closed loop, the number of left and right turns differ by 6, summing over both orientations gives loops a fugacity $n=2\cos 6{\pi\over 12}=0$ as requested.  Other examples of complex Boltzmann weights occur in the Ising model in an imaginary magnetic field, where the Yang-Lee edge singularity describes critical points of hard objects with negative fugacity~\cite{CardyYL}. 

\begin{figure}
\begin{center}
\psfig{file=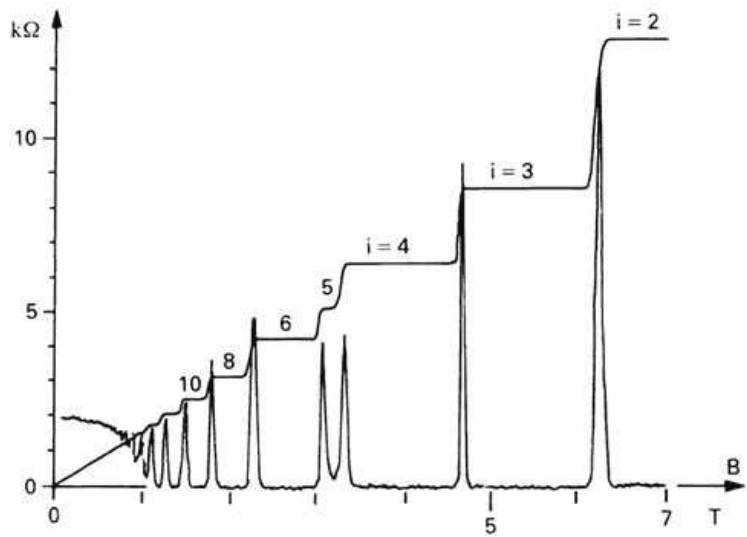,width=4in}
\end{center}
\caption{Plateaus for the Hall resistance and peaks of the Ohmic resistance  in the integer quantum Hall effect. Neighboring  values of $i$ are separated by a quantum critical point, whose properties are described by a $c=0$ two-dimensional (2D) CFT. The figure is from ${\rm {\tt http://www.nobelprize.org/nobel\_prizes/physics/laureates/1998/press.html}}$.}
\label{aba:fig1}
\end{figure}

Apart from geometrical problems, another situation where non-unitary CFTs are crucial is the description of critical points in non-interacting $2+1$ dimensional disordered systems, such as the transition between plateaux in the integer quantum Hall effect~\cite{Klitzing, Tsui} (see Fig~\ref{aba:fig1}). There, transport properties after average over disorder can be expressed in terms of a two-dimensional sigma model~\cite{Pruisken} on a super-coset~\cite{Efetov} of the type 
\begin{equation}
{U(1,1|2)\over U(1|1)\times U(1|1)}
\end{equation}
with topological angle $\theta=\pi$. This sigma model is naturally  non-unitary -- a physical consequence  of averaging over disorder (which could also be done using replicas) --  because of general supergroup properties (see below). It is believed to flow to a strongly interacting CFT with central charge $c=0$~\cite{Zirnbauer}, whose (unknown) exponents describe the plateau transition. 
Another similar disordered problem of non-interacting fermions described by a LCFT is given by the spin quantum Hall effect~\cite{SQHE}. It turns out that a subset of operators in this theory can be described in terms of a sigma model on a compact super-coset
\begin{equation}
\mathbb{CP}^{1|1}={U(2|1)\over U(1)\times U(1|1)}
\end{equation}
which is closely related to the classical percolation problem~\cite{SQHE,RS1}.

More generally, we note that there have been suggestions that the canonical formulation of quantum mechanics itself be relaxed to allow for non-Hermitian, PT symmetric Hamiltonians~\cite{Bender}. Such hamiltonians might, for instance, describe driven open systems, for which, once again, non-unitarity is natural.

Now, non-unitarity might be expected to be only a minor nuisance. After all, plenty of manageable (rational) non-unitary CFTs are known, whose central charges are given by a formula generalizing (\ref{cunit}):
\begin{equation}
c=1-{6(p-q)^2\over pq} ,\quad p\wedge q=1\label{genc}
\end{equation}
For instance the case $p=5$ and $q=2$ describes the Yang-Lee singularity mentioned earlier~\cite{CardyYL}. The associated CFT admits a negative central charge $c=-{22\over 5}$ and a negative conformal weight $h=-{1\over 5}$. In terms of the Virasoro algebra 
\begin{equation}
\left[L_n,L_m\right]=(n-m)L_{n+m}+{c\over 12}(n^3-n)\delta_{n+m}
\end{equation}
this means that the quadratic form defined by $L_n^\dagger\equiv L_{-n}$ is not positive definite. For instance, the state associated with the stress-energy tensor has a negative norm-square
\begin{equation}
\langle T|T\rangle=\langle 0|L_2L_{-2}|0\rangle={c\over 2}=-{11\over 5}.
\end{equation}
Nevertheless, most properties of this theory can be handled like those for the unitary series~\eqref{cunit}, the only difference being in some unphysical signs.

It turns out however that for the cases  of more direct physical interest  the consequences of non-unitarity are considerably more important. First, many physically reasonable results stop working -- an example of this is the failure of the Mermin--Wagner theorem for two dimensional models with continuous  (super) symmetry \cite{Forests}. More technically maybe, giving up hermiticity means that Hamiltonians or transfer matrices are not necessarily diagonalizable any longer. Combining this feature with criticality in $1+1$ dimensions leads to the possibility of Jordan cells for the dilatation operator $L_0$, and therefore, to profound physical and mathematical modifications of conformal invariance, giving rise to what is called Logarithmic Conformal Field Theory (LCFT)~\cite{Gurarie,Flohrreview,Gaberdielreview} --  a proverbially intricate subject, where progress has been slow for many years. Things have improved recently, thanks in part to a rather down to earth, lattice approach (combined with  progress in  the theory of associative algebras) that we review here.  To explain what happens, we start by  making a detour through representation theory of supergroups.

\section{Indecomposability: the Lie superalgebra $g\ell(1|1)$ and its representations}

\subsection{Defining relations}

The Lie superalgebra $g\ell(1|1)$ is generated by two bosonic elements 
$E,N$ and two fermionic generators $\Psi^\pm$ such that $E$ is central 
and the other generators obey
\begin{equation}
  [N,\Psi^\pm] \ = \ \pm \Psi^\pm \ \ \ \mbox{and} \ \ \ \ 
    \{ \Psi^-,\Psi^+\} \ = \ E. 
\end{equation}
The even subalgebra is thus given by $g\ell(1)\oplus g\ell(1)$. 
Let us also fix the following Casimir element $C$ 
$$ C \ = \ (2N-1) E + 2 \Psi^-\Psi^+ . $$
The choice of $C$ is not unique since we could add any 
function of the central element $E$. This has interesting consequences in field theory.

Finally we recall the definition of the supertrace $\hbox{Str}(\cdot)=\hbox{Tr}((-1)^F\cdot)$. The superdimension is the supertrace of the identity, {\it i.e.}, the number of bosons minus the number of fermions. The superdimension of  $g\ell(1|1)$ is zero. The representation theory can be summarized quite easily  -- see \cite{Gotz} for details and references.

\subsection{Irreducible representations}

To begin with, we list the 
irreducible representations which fall into two different 
series. There is one series of two-dimensional representations 
$\langle e,n\rangle$ which is labeled by pairs $e,n$ with $e\neq 0$ and 
$n \in \mathbb{R}$. In these representations, the generators take the 
form $E= e {\bf 1}_2$ and 
$$  N \ = \ \left(\begin{matrix} n-1 & 0 \\ 0 & n \end{matrix}\right) 
\ \ , \ \ 
 \Psi^+ \ = \ \left(\begin{matrix} 0 & 0 \\ e & 0 \end{matrix}\right) 
\ \ , \ \ 
 \Psi^- \ = \ \left(\begin{matrix} 0 & 1 \\ 0 & 0 \end{matrix}\right) 
\ \ . 
$$   
These representations are the typical representations (long 
multiplets). In addition, there is one series of 
atypical representations $\langle n\rangle$ (short multiplets). These are 
1-dimensional and parametrized by the value $n \in \mathbb{R}$ of $N$. 
All other generators vanish.

\subsection{Indecomposability in tensor products}

Having seen all the irreducible representations $\langle e,n\rangle$ and
$\langle n\rangle $ of $g\ell(1|1)$, 
our next task is to compute tensor products of typical 
representations $\langle e_1,n_1\rangle$ and $\langle e_2,n_2\rangle$ using a basis $\lbrace |0\rangle, |1\rangle \rbrace$. 
Here, we emphasize that we deal with graded tensor products, that is, when we pass a fermionic operator through a fermionic state, we generate an additional minus sign. We will take the convention that $\Ket{0}$ is bosonic and $|1\rangle$ is fermionic for the time being. It is of course possible to switch the $\mathbb{Z}_2$ grading and decide that $\Ket{0}$ is fermionic, {\it etc}.
As long
as $e_1+e_2\neq 0$, the tensor product is easily seen to 
decompose into a sum of two typicals, 
\begin{equation}
 \langle e_1,n_1\rangle \otimes \langle e_2,n_2\rangle \ = \ \langle e_1+e_2,n_1+n_2-1\rangle
  \oplus \langle e_1+e_2,n_1+n_2\rangle.
  \end{equation}
But when $e_1+e_2=0$ we obtain a 4-dimensional representation 
that cannot be decomposed into a direct sum of smaller 
subrepresentations! The representation matrices of these
4-dimensional indecomposables ${\cal P}_{n}$ read as 
follows ($n\equiv n_1+n_2-1$, $E=0$)
$$ N \ = \ \left(\begin{matrix} n-1 & 0 & 0 & 0 \\ 0 & n & 0 & 0 
 \\ 0 & 0 & n & 0 \\ 0 & 0 & 0 & n+1 \end{matrix}\right)  
\ \ , \ \ 
 \Psi^+ \ = \ \left(\begin{matrix} 0 & 0 & 0 & 0  \\ 
             -1 & 0 & 0 & 0 \\ 1 & 0 & 0 & 0 \\
                  0 & 1 & 1 & 0 \end{matrix} \right)
\ \ , \ \ 
 \Psi^- \ = \ \left(\begin{matrix} 0 & 1 & 1 & 0  \\ 
 0 & 0  & 0 & 1 \\ 0 & 0 & 0 & -1 \\ 0 & 0 & 0 & 0 
\end{matrix}\right) \ \ . 
$$   
It is useful to picture the structure 
of indecomposables. The form of $N$ tells us that ${\cal P}_{n}$ 
is composed from the atypical irreducibles $\langle n-1\rangle$, $2\langle n\rangle$, and  
$\langle n+1\rangle$. The action of $\Psi^\pm$ relates these four 
representations as follows
\begin{equation}
 {\cal P}_n \qquad
  \langle n\rangle\ \lra\ \langle n+1\rangle\oplus\langle n-1\rangle\ 
\lra\ \langle n\rangle\ \ .
\end{equation}
or, more explicitly,
\begin{align}
   \xymatrix@C=5pt@R=26pt@M=2pt{%
    &&\\
    &{\cal P}_n: &\\
    &&
 }     
&  \xymatrix@C=9pt@R=24pt@M=3pt@W=2pt{%
    &&{\langle n\rangle }\ar[dl]_{\Psi^-}\ar[dr]^{\Psi^+}&\\
    &\langle n-1\rangle \ar[dr]_{\Psi^+}&&\langle n+1\rangle \ar[dl]^{\Psi^-}\\
    &&\langle n\rangle &
 }
\end{align}   
We refer to the structure of indecomposable modules in terms of simple modules and mappings between them as subquotient structure. The  Casimir element $C$ in 
the representations ${\cal P}_n$ maps the subspace $\langle n\rangle$ on the top onto the $\langle n\rangle$ 
on the bottom of the above diagram and is zero otherwise.
This means that $C$ cannot be diagonalized in ${\cal P}_n$. We 
shall return to this observation later on.

%

\section{Indecomposability in CFT}

\subsection{The GL$(1|1)$ WZW model}

Imagine now building the WZW model with $g\ell(1|1)$ symmetry~\cite{RozSal}. Naturally, there will be primary fields  $\Phi_{\langle e,n\rangle}$ associated with the typical representations $\langle e,n\rangle$, and their conjugates associated with $\langle -e,1-n\rangle$. Now the fact that the corresponding tensor product in $g\ell(1|1)$ is not fully reducible translates into a strange behavior of the operator product expansion (OPE), where the `merging' of the two representations on the right hand side of the tensor product translates into logarithmic terms~\cite{SchomSal}.  The details of this logarithmic OPE are not essential here -- the point, rather, is the presence of the logarithm, justifying the name of Logarithmic Conformal Field Theory (LCFT). Alternatively, observe that the Sugawara stress tensor's zero mode
\begin{eqnarray}
L_0&=&{1\over 2k}\left(2N_0E_0-E_0+2\Psi_0^-\Psi_0^++{1\over k}E_0^2\right)\nonumber\\
&+&{1\over k}\sum_{m\geq 0} \left(E_{-m}N_m+N_{-m}E_m+\Psi_{-m}^-\Psi_m^+-\Psi_{-m}^+\Psi_m^-+{1\over k}E_{-m}E_m\right)
\end{eqnarray}
has a rank-two Jordan cell in the Kac Moody representations based on ${\cal P}_n$, and is not diagonalizable. This is the behavior characteristic of LCFT, and is possible because $L_0$ is not hermitian.

\subsection{$c=0$ catastrophe}
\label{subsecCatastrophe}

While the initial example of the $g\ell(1|1)$ WZW theory~\cite{RozSal} was perceived as a bit marginal, V.~Gurarie pointed out soon after that indecomposability and logarithms are almost unavoidable at $c=0$. The argument is as follows. Based on general conformal invariance arguments, one can show that, whenever there is a single field with conformal weights $h=2$, $\bar{h}=0$, the OPE of a primary field with itself should take the form
\begin{equation}
\Phi(z)\Phi(0)\approx a_\Phi z^{-2h_\Phi}\left(1+{2h_\Phi\over c}z^2 T(z)+\ldots\right).
\end{equation}
If one now imagines reaching $c=0$ through a limit process within a family of CFTs -- for instance by taking the $n\rightarrow 0$ limit of the $O(n)$ model for self-avoiding walks, or by calculating the average free energy of a disordered system in a replica approach -- there will occur a $c=0$ `catastrophe' \cite{Gurarie} because of the vanishing denominator. In general, this is not acceptable (although the issue is a little  subtle), so the divergence must be canceled by another term coming from a field whose dimensions are not those of the stress tensor generically, but become $h=2,\bar{h}=0$ right at $c=0$. The OPE then reads
\begin{equation}
\Phi(z)\Phi(0)\approx a_\Phi z^{-2h_\Phi}\left(1+{2h_\Phi\over b}z^2 \left[t(0)+\ln z \ T(0)\right]+\ldots\right)
\end{equation}
where $t(z)$ is called the logarithmic partner of $T(z)$. One can also show that $L_0$ then has a Jordan cell of rank two, and 
\begin{equation}
L_0|T\rangle =2|T\rangle,~~L_0|t\rangle=2|t\rangle +|T\rangle
\end{equation}
together with 
\begin{eqnarray}
\langle T(z)T(0)\rangle&=&0,\nonumber\\
\langle T(z)t(0) \rangle&=&{b\over z^4},\nonumber\\
\langle t(z)t(0)\rangle&=&{-2b\ln z+a\over z^4}.
\end{eqnarray}
Note that $T$ is a null state, but it is not decoupled. Note also that the equations involve a parameter $b$ (called logarithmic coupling) whose value is {\it a priori} undetermined. Finally, 
note that $t$ is not a true scaling field, since it `mixes' with $T$ under a scale transformation. But it is important to stress that the logarithm occurs right at the fixed point of the renormalization group: this is very different from logarithmic corrections that may appear in models with marginally irrelevant operators, such as the XY model at the Kosterlitz-Thouless point~\cite{Eggert}.  For more details on the transformation rules of $T(z)$ and $t(z)$, see the reviews~\cite{Flohrreview,Gaberdielreview}.

An important question that arises when comparing this section and the previous one is the existence of `hidden'  symmetries in $c=0$ theories explaining the degeneracy at conformal weight $h=2$ necessary to cure the $c=0$ catastrophe. This is  discussed  in J. Cardy's review in this volume (from the point of view of replicas)~\cite{CardyThis}, and below using our lattice approach in section~\ref{sec:fusion}. See also 
\cite{CardyLCFT,Gurariecthm}.

\section{Rationale for a lattice approach}

The spectacular progress in our understanding of ordinary CFTs occurred largely due to a better  understanding of the representation theory of the Virasoro algebra. The decoupling of zero norm states of given conformal weight in particular leads to the Kac table of (not necessarily unitary) models  (\ref{genc}), differential equations for the correlators of the corresponding primary fields, {\it etc}. In the case of LCFTs however - such as those at $c=0$ describing polymers and percolation - decoupling such states voids the theory of most or all its physical content, and they must absolutely be kept. The problem is that there are then very  few tools left to constrain a priori the content of the LCFT.  The general approaches which were so successful in the case of ordinary CFTs simply lead nowhere, and despite years of effort and many interesting partial results (see {\it e.g.}~\cite{FlohrAML,[FGST3],MathieuRidout,GabRun,GabRunW2,GabRunW}), very little solid information about even the simplest physically relevant LCFTs was available. 

 One avenue for recent progress has been to try to gain knowledge about the algebraic features of the continuum limit by studying those of lattice regularizations. This seems a priori a bit hopeless, as precisely much of the power of CFT has come originally from the fact that it deals directly with the continuum theory, where more, infinite dimensional symmetries are available. However, as time went on, it was realized that most of the power of CFTs arose from algebraic structures that were present also -- albeit in some finite-dimensional form -- on the lattice, such as quantum groups~\cite{PasquierSaleur,QG-book} and their centralizers~\cite{M1}. Note that quantum groups naturally arise in CFT~\cite{QG-book} as well and might be very useful in getting some information on LCFTs~\cite{[FGST2],[FGST3],[FGST4],[GT],[BGT]} (see also the article~\cite{ST2} of A.~Semikhatov and I.~Tipunin in this volume). Hence, trying to apply this lesson to understand LCFTs is in fact not so counterintuitive. Another reason for trying to be as concrete as possible is that the landscape of LCFTs seems unbelievably complicated to say the least: it is not enough to build `one' consistent LCFT at $c=0$ to solve the problem -- there is most likely an infinity of such theories, and  a lot of additional information  must be provided to ensure correct identification. 

 As we discussed briefly  in the introduction, we are especially interested in LCFTs which are  fixed points of  interacting, non-unitary, field theories with well defined local actions, such as the super-projective sigma models at topological angle $\theta=\pi$, {\it etc}.  If such LCFTs exist, it is reasonable to expect that they must also admit some lattice regularizations with local degrees of freedom, that is, that their properties can be studied by considering models defined on large, but finite, lattices, and exploring their continuum or scaling limits. The point is to look, now, at \textit{finite-dimensional} algebras describing the dynamics in such finite-lattice models, which are typically quantum spin-chains with {\sl local} interaction. For most of the physically relevant cases these algebras are  the Temperley--Lieb algebras and its boundary extensions which will be defined below.
 It is important that the representation theory of these \text{lattice} algebras (representations, fusion, etc) is well under control. The  hope beyond the lattice approach -- which, as we will see, is warranted by experience -- is that one can  study all physically relevant indecomposable Virasoro modules and their fusion rules (and  probably more) by simply defining them as scaling limits of spin-chains modules. 
 
 Before we describe our lattice regularization approach in details, let us also mention other related works that also use lattice models in order to probe the complicated structure of LCFTs. Whereas our analysis focuses mainly on quantum spin chains, it is also interesting to study thoroughly loop models that provide indecomposable representations of the Temperley-Lieb algebra~\cite{AubinLoop}. See also the contribution from Morin-Duchesne and Saint-Aubin in this special issue~\cite{AubinLoopThis}. Another lattice approach, with a special emphasis on fusion, was developed by Pearce, Rasmussen and Zuber in~\cite{PRZ} and pushed further in {\it e.g.}~\cite{RP1,RP2,PRV}.

\section{Quantum spin chains, Temperley-Lieb algebra and indecomposability}


In this preliminary section, we begin by introducing the ``simplest'' lattice models relevant for physics.
We will mostly focus on the $\LQG$-invariant open XXZ spin chain~\cite{PasquierSaleur} but we also describe
how supersymmetric spin chains can be dealt with in the same way.  We show how this relates to representations
of the  Temperley--Lieb algebra (see~\cite{GV,RidoutSaintAubin} for recent reviews,  and the  book~\cite{M0} for crucial pioneer work), for which we provide a short review of the representation theory.

\subsection{Temperley-Lieb algebra and XXZ spin chain}

The simplest class of lattice models that we will study throughout this paper provides representations of the 
so-called Temperley-Lieb (TL) algebra $\TLq{N}$ defined on $N$ sites. It is generated by $e_i$'s, with $1\leq i\leq N-1$, and has the defining relations
\ba 
\left[ e_i , e_j \right] &=&0, \qquad\left|i-j \right| \geq 2, \label{TLdef}\\
e_i ^2 &=& \fug e_i,\\
e_i e_{i \pm 1} e_i &=& e_i,\label{TLdef-e}
\ea
with
\be
\displaystyle{ \fug=\q+\q^{-1}},
\ee
and $\q$ is a parameter which can in general take any complex value, but all physically interesting cases require $\q$ to be a root of unity. In particular, dense polymers correspond to $\q=e^{i\pi/2}$ and percolation corresponds to $\q=e^{i\pi/3}$.

It is well-known that the TL algebra can be thought of as an algebra of diagrams~\cite{M0}. Using the notation
$$
\begin{pspicture}(0.,0.)(1.0,1.0)
	\psline[](0.,0.)(0.,0.6)
	\psline[](0.5,0.)(0.5,0.6)
	\rput{0}(-0.7,0.25){$e_i=$}
\end{pspicture}\hdots \ \ \
\begin{pspicture}(0.,0.)(1.0,1.0)
	\psellipticarc{-}(0.25,0.)(0.25,-0.25){180}{0}
	\psellipticarc{-}(0.25,0.6)(0.25,0.25){180}{0}
	\rput{0}(0.,-0.3){$^i$}
	\rput{0}(0.5,-0.3){$^{i+1}$}
\end{pspicture} \hdots \ \ \
\begin{pspicture}(0.,0.)(1.0,1.0)
	\psline[](0.,0.)(0.,0.6)
	\psline[](0.5,0.)(0.5,0.6)
\end{pspicture},
$$
eqs.~\eqref{TLdef}-\eqref{TLdef-e} can now be interpreted geometrically. The
composition law corresponding to stacking the diagrams of the $e_i$'s vertically,
where it is assumed that every closed loop carries a weight $\fug$, henceforth called the fugacity of a loop.

 \begin{figure}[t!]
\begin{center}
{\footnotesize
\begin{pspicture}(0,0)(7,9)
\psset{xunit=8mm,yunit=8mm}


\psellipticarc{-}(1.75,6.0)(0.5,-0.5){180}{0}
\psellipticarc{-}(1.75,7.5)(0.5,0.5){0}{359}
\psellipticarc{-}(1.75,9.0)(0.5,0.5){180}{0}

  \rput[Bc](3.5,7.5){$=$}
  \rput[Bc](4,7.5){$\fug$}
  
  \psellipticarc{-}(5,6.75)(0.5,-0.5){180}{0}
  \psellipticarc{-}(5,8.25)(0.5,0.5){180}{0}
  
  \psline{-}(4.5,6)(4.5,6.75)
  \psline{-}(5.5,6)(5.5,6.75)
  
  \psline{-}(4.5,8.25)(4.5,9)
  \psline{-}(5.5,8.25)(5.5,9)
  

\psellipticarc{-}(1.0,0.)(0.5,-0.5){180}{0}
\psellipticarc{-}(1.0,1.5)(0.5,-0.5){0}{180}
\psellipticarc{-}(2.0,1.5)(0.5,-0.5){180}{0}
\psellipticarc{-}(2.0,3.0)(0.5,-0.5){0}{180}
\psellipticarc{-}(1.0,3.0)(0.5,-0.5){180}{0}
\psellipticarc{-}(1.0,4.5)(0.5,-0.5){0}{180}

\psline{-}(2.5,0.0)(2.5,1.5)
\psline{-}(2.5,3.0)(2.5,4.5)
\psline{-}(0.5,1.5)(0.5,3.0)

  \rput[Bc](3.5,2.25){$=$}
  
  \psellipticarc{-}(5.0,1.5)(0.5,-0.5){180}{0}
  \psellipticarc{-}(5.0,3.0)(0.5,-0.5){0}{180}
  \psline{-}(4.5,0)(4.5,1.5)
  \psline{-}(5.5,0)(5.5,1.5)
  \psline{-}(4.5,3)(4.5,4.5)
  \psline{-}(5.5,3)(5.5,4.5)
  \psline{-}(6.5,0)(6.5,4.5)
  

\end{pspicture}
}
\end{center}
\caption{Interpretation of the Temperley-Lieb algebra defining relations in terms of diagrams.}
\label{aba:figTL}
\end{figure}
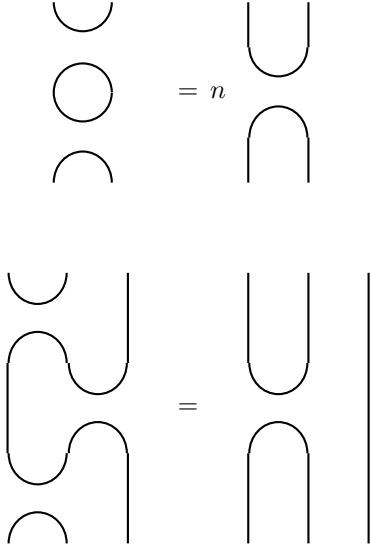

We now consider the Hamiltonian limit of the six-vertex 2D lattice model with open boundary conditions. This limit is described
by (1+1)D system or a spin-$\frac{1}{2}$ quantum chain with Heisenberg-like interactions. Each local interaction term 
is given by so-called XXZ representations of the TL algebra.
The representation of $\TLq{N}$ on the spin-chain space $\mathcal{H}_{N}=(\mathbb{C}^2)^{\otimes N}$
is given by 
\begin{equation}\label{TL-XXZ}
\displaystyle e_i = \ffrac{\q + \q^{-1}}{4} - \frac{1}{2} \left(
\sigma^x_i \sigma^x_{i+1} + \sigma^y_i \sigma^y_{i+1} + \ffrac{\q +
\q^{-1}}{2} \sigma^z_i \sigma^z_{i+1} \right) - \ffrac{\q - \q^{-1}}{4}
\left( \sigma^z_i - \sigma^z_{i+1} \right)
\end{equation}
in terms of the usual Pauli matrices.

The Temperley-Lieb generators $e_i$ can then be thought of as the Hamiltonian densities
of
the XXZ Hamiltonian~\cite{PasquierSaleur} with additional boundary terms
\be\label{XXZ_H}
\displaystyle H = \frac{1}{2} \sum_{i=1}^{N-1} \left( \sigma^x_i
\sigma^x_{i+1} + \sigma^y_i \sigma^y_{i+1} + \ffrac{\q + \q^{-1}}{2}
\sigma^z_i \sigma^z_{i+1} \right) + \ffrac{\q - \q^{-1}}{4} \left(
\sigma^z_1 - \sigma^z_{N} \right).
\ee
Up to an irrelevant constant term, we thus have 
\be   H = - \sum_{i=1}^{N-1} e_i.
\ee

\subsection{Lattice symmetries: quantum groups and bimodules}
It is important to discuss the symmetries of the lattice models. Recall that the usual Heisenberg XXX spin chain ($\q=1$ in~\eqref{XXZ_H}) is defined by its  Hamiltonian $H$ acting in the vector space $\mathcal{H}_{N}$
\begin{equation}
H=\sum_i \vec{S}_i\cdot\vec{S}_{i+1}, \qquad \mathcal{H}_{N}=\square^{\otimes N}
\end{equation}
where $\square = \mathbb{C}^2$ denotes the fundamental representation of $s\ell(2)$. It is an antiferromagnetic chain, and accordingly its nearest neighbor coupling $\vec{S}_i\cdot\vec{S}_{i+1}$, projects neighbor pairs of spins onto the singlet. The continuum limit is well-known to be described by the $O(3)$ sigma model at $\theta=\pi$ which flows to the level-1 $SU(2)$ WZW theory at low energy. 

\begin{figure}
\begin{center}
\includegraphics[scale=0.9]{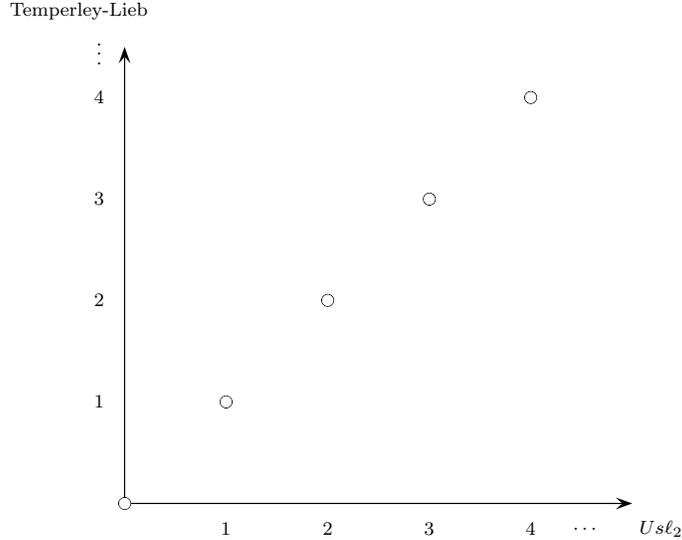}
%
%
%
%
%
%
%
%
%
%
%
\end{center}
\caption{Bimodule for the antiferromagnetic Heisenberg or XXX spin chain and $N$ even. Integers that appear along the horizontal and vertical axis correspond to $j$, the spin for $s\ell(2)$, and half the number of though-lines for $\TLq{N}$. }
\label{aba:fig4}
\end{figure}

For the Heisenberg or XXX spin chain, there are two natural algebras to consider. One is the symmetry algebra $s\ell(2)$ generated by
$S^{\pm}$ and $S^z$ operators satisfying the usual relations
\be\label{sl-rel}
[S^+,S^-] = 2 S^z,\qquad [S^z,S^{\pm}]=\pm S^{\pm}.
\ee
Recall that the $s\ell(2)$ action is defined by the iterated comultiplication $\Delta(S^a)= \one\otimes S^a + S^a\otimes\one$ which gives on the full tensor product the action
\be
S^{a}v = \sum_{i=1}^{N} S^a_i v, \qquad v\in\Hilb_N,\quad a=\pm,z,
\ee
where $S^a_i$ is the action on the $i$th fundamental representation.
The other symmetry algebra is generated by the local hamiltonian densities $\vec{S}_i\cdot\vec{S}_{i+1}$. This algebra actually coincides  with (a quotient of) the group algebra of the permutation group, which is nothing in this case but the Temperley--Lieb algebra for the value $\fug=2$ of the fugacity parameter $\fug$. The actions of the two algebras of course commute -- the symmetry commutes not only with the Heisenberg Hamiltonian but also with all its densities. Moreover, the $s\ell(2)$ algebra is the {\sl full} symmetry algebra of the densities, or more technically, it is the \textit{centralizer} of the  Temperley--Lieb algebra with $\q=1$. Any operator commuting with the generators $\vec{S}_i\cdot\vec{S}_{i+1}$ belongs to the representation of $s\ell(2)$, and actually vice versa. So, both algebras are mutual centralizers. What this really means is that we can decompose the Hilbert space $\Hilb_{N}$ in terms of $s\ell(2)$ representations of (integer, if we restrict to chains of even length) spin $j$; then, the vector space of all highest-weight states of a given spin $j$ provides an irreducible representation of the permutation group or the TL algebra. Its dimension is obviously the multiplicity of the spin $j$ representation of $s\ell(2)$ and it is given by the numbers
\begin{equation}
\dd_j=\binom{N}{N/2+j} -
\binom{N}{N/2+j+1},\qquad
\text{and we set}\; \dd_j=0\; \text{for}\; 2j>N.
\end{equation}
This representation is irreducible by construction where we  used the mutual centralizers property. The full Hilbert space of states can thus be considered not just a representation (or equivalently a module) for one of the algebras, but rather a {\sl bi}-module for both algebras simultaneously.
In other words, the space of states $\Hilb_N$, as a semi-simple bi-module over the  pair of commuting algebras, can be decomposed as
\begin{equation}\label{Hilb-decomp-gen-bimod}
\displaystyle \Hilb_{N} \cong \bigoplus_{j=(N\modd2)/2}^{N/2} \StTL{j}[N]\otimes\modWeylj{j},
\end{equation}
where the first algebra generated by the densities $\vec{S}_i\cdot\vec{S}_{i+1}$ acts on the left tensorands denoted by $\StTL{j}[N]$,
 while the second algebra which is $s\ell(2)$ acts on the right components which are spin $j$ modules denoted by $\modWeylj{j}$, and these $s\ell(2)$-modules do not depend on $N$.
 Finally, the resulting bimodule can be represented graphically as in Fig.~\ref{aba:fig4}, where each open dot represents a simple (irreducible) module for both algebras.

The  Hamiltonian~\eqref{XXZ_H} of the XXZ model now generalizes this usual Heisenberg (or XXX) model to a spin chain with \text{quantum-group} $\LQG$ symmetry. This symmetry is generated by the $S^{\pm}$ and $S^z$ operators that now satisfy the quantum-group relations
\be\label{qsl-rel}
[S^+,S^-] = \frac{\q^{2S^z}-\q^{-2S^z}}{\q-\q^{-1}},\qquad [S^z,S^{\pm}]=\pm S^{\pm},
\ee
which are just $\q$-deformed versions of the usual relations~\eqref{sl-rel}. The action of $\LQG$ on the spin-chain is obtained again using iteratively the (deformed) comultiplication 
\be
\Delta(S^a)= \q^{S^z}\otimes S^a + S^a\otimes\q^{-S^z},\qquad \Delta(\q^{\pm S^z}) = \q^{\pm S^z}\otimes \q^{\pm S^z}.
\ee
An explicit action can be found in~\cite{PasquierSaleur}, for example.

When $\q$ is generic, {\it i.e.} not a root of unity, the Hilbert space of the Hamiltonian densities~\eqref{TL-XXZ} nicely decomposes
onto the irreducible $\TLq{N}$ representations of the same dimensions $\dd_j$ as before, so we will denote them by $\StTL{j}[N]$ again,
\begin{equation}\label{Hilb-decomp-gen}
\displaystyle \Hilb_{N} |_{\rule{0pt}{7.5pt}%
\TLq{N}} \cong \bigoplus_{j=(N\modd2)/2}^{N/2} (2j+1) \StTL{j}[N],
\end{equation}
where the degeneracies $2j+1$ correspond to the dimension of the spin $j$ representations or
so-called Weyl modules (which are also generically irreducible) over the
symmetry algebra for $\TLq{N}$, which is 
the quantum group $\LQG$.
We can thus consider the space $\Hilb_N$ again as a semi-simple bi-module over the  pair of commuting algebras $\TLq{N}$ and $\LQG$, and it has the same decomposition as in~\eqref{Hilb-decomp-gen-bimod}.

 Things become more intricate when $\q$ is a root of unity, which corresponds
to most of the physically relevant cases. We shall denote $\q=\mathrm{e}^{i\pi /p}$ in this case,
and we will use the following denominations, borrowed from the Potts model terminology, 
for the several physically relevant cases:
 dense polymers ($p=2$), percolation ($p=3$), Ising model ($p=4$), {\it etc}.
In these cases, the algebra $\TLq{N}$ is non-semisimple and the decomposition~\eqref{Hilb-decomp-gen} is no longer true. We will describe the structure of the XXZ spin-chain at these roots of unity cases after a short detour around the representation theory of the TL algebra.

\subsection{Superspin chains}

Another natural way to construct spin chain representations of the TL algebra
is given by supersymmetric (SUSY) spin chain~\cite{ReadSaleur01}. 
We construct these spins chains in the following way:
each site carries a $\mathbb{Z}_2$ graded vector space of dimension $n+m|m$,
that is, a bosonic ({\it resp.} fermionic) space of dimension $n+m$ ({\it resp.} $m$). 
We choose these vector spaces to be the fundamental $\square$ of the Lie superalgebra 
$g\ell(n+m|m)$ for $i$ odd
and the dual $\bar{\square}$ for $i$ even -- recall that $i$ labels the spin positions. The Hamiltonian $H=- \sum_i e_i$
then acts on the graded tensor product $\mathcal{H} = (\square \otimes \bar{\square})^{\otimes N}$. 
The TL generators are defined (up to a multiplicative constant)
as projectors onto the singlet in the tensor products $\square \otimes \bar{\square}$
and $ \bar{\square} \otimes \square$. 
These superspin chains describe the strong coupling region of a non-linear $\sigma$-model~\cite{ReadSaleur01} on the complex projective superspace 
\begin{equation}
\mathbb{CP}^{n+m-1|m} = \mathrm{U}(m+n|m) / (\mathrm{U}(1) \times \mathrm{U}(m+n-1|m)),
\end{equation}
at topological angle $\theta = \pi$. 

In this paper, we will mostly focus on the $\LQG$-invariant XXZ spin chain for pedagogical reasons, but
most of our results can also be understood in terms of these supersymmetric spin chains~\cite{RS2}.
It is worth mentioning that this supersymmetric formulation turns out to be particularly convenient when 
dealing with {\it periodic} systems, as the $\LQG$ symmetry of the XXZ spin chain is lost in that case (see {\it e.g.}~\cite{GRS1,GRS2,GRS3, GRSV}).

\subsection{A short review of the representation theory of $\TLq{N}$}

\label{subsecRepTheoryTL}

 It is well known~\cite{M0,M1} that when  $\q$ is generic,
{\it i.e.} not a root of unity, the representation theory of $\TLq{N}$ is  semi-simple.
All simple modules or irreducible representations in this case are described geometrically by
so-called standard modules. 
For $j$ (half-)integer such that $0\leq j\leq N/2$ and on $N$ sites, we define a standard module
$\StTL{j}[N]$ with $2j$ through-lines (also called ``strings'') as the span of link diagrams -- all possible nested configurations of $(\frac{N}{2}-j)$ arcs, like \
$\psset{xunit=2mm,yunit=2mm}
\begin{pspicture}(0,0)(7,1)
 \psellipticarc[linecolor=black,linewidth=1.0pt]{-}(0.5,1.0)(0.5,1.42){180}{360}
 \psellipticarc[linecolor=black,linewidth=1.0pt]{-}(2.5,1.0)(0.5,1.42){180}{360}
 \psellipticarc[linecolor=black,linewidth=1.0pt]{-}(5.5,1.0)(0.5,0.71){180}{360}
 \psellipticarc[linecolor=black,linewidth=1.0pt]{-}(5.5,1.0)(1.5,1.42){180}{360}
 \\
\end{pspicture}$\;.
Through-lines are denoted by a vertical line $\thl$ and are not allowed to intersect any arc.
The action of the generators on these modules is again interpreted as stacking
the various diagrams with the additional rule that contracting any pair of strings results in zero. 
The dimension of these standard modules reads
\begin{equation}\label{eqDimStdTL}
\dim(\StTL{j}[N])\equiv \dd_j = \binom{N}{N/2+j} -
\binom{N}{N/2+j+1},\qquad
\text{and we set}\; \dd_j=0\; \text{for}\; 2j>N.
\end{equation}
We stress that $\dd_j$ does not depend on $\q$.
Note that $j$ must be half integer when $N$ is odd.
For $N=4$ for instance, there are four standard modules with basis
\begin{align}
\StTL{0}[4]&=
\{ \
\psset{xunit=2mm,yunit=2mm}
\begin{pspicture}(0,0)(3,1)
 \psellipticarc[linecolor=black,linewidth=1.0pt]{-}(1.5,1.0)(1.5,1.42){180}{360}
 \psellipticarc[linecolor=black,linewidth=1.0pt]{-}(1.5,1.0)(0.5,0.71){180}{360}
\end{pspicture}
,\, \psset{xunit=2mm,yunit=2mm}
\begin{pspicture}(0,0)(3,1)
 \psellipticarc[linecolor=black,linewidth=1.0pt]{-}(0.5,1.0)(0.5,1.42){180}{360}
 \psellipticarc[linecolor=black,linewidth=1.0pt]{-}(2.5,1.0)(0.5,1.42){180}{360}
\end{pspicture} \
\},\\
\StTL{1}[4] &=
\{ \ 
\psset{xunit=2mm,yunit=2mm}
\begin{pspicture}(0,0)(3,1)
\psline[linewidth=1.0pt](0,-0.5)(0,1)
\psellipticarc[linecolor=black,linewidth=1.0pt]{-}(1.5,1.0)(0.5,1.42){180}{360}
\psline[linewidth=1.0pt](3,-0.5)(3,1)
\end{pspicture}
\ ,\, \psset{xunit=2mm,yunit=2mm}
\begin{pspicture}(0,0)(3,1)
 \psellipticarc[linecolor=black,linewidth=1.0pt]{-}(0.5,1.0)(0.5,1.42){180}{360}
\psline[linewidth=1.0pt](2,-0.5)(2,1)
\psline[linewidth=1.0pt](3,-0.5)(3,1)
\end{pspicture}
\ , \, \psset{xunit=2mm,yunit=2mm}
\begin{pspicture}(0,0)(3,1)
\psline[linewidth=1.0pt](0,-0.5)(0,1)
\psline[linewidth=1.0pt](1,-0.5)(1,1)
\psellipticarc[linecolor=black,linewidth=1.0pt]{-}(2.5,1.0)(0.5,1.42){180}{360}
\end{pspicture} \
\}
,\\
\StTL{2}[4] &=
\{ \
\psset{xunit=2mm,yunit=2mm}
\begin{pspicture}(0,0)(3,1)
\psline[linewidth=1.0pt](0,-0.5)(0,1)
\psline[linewidth=1.0pt](1,-0.5)(1,1)
\psline[linewidth=1.0pt](2,-0.5)(2,1)
\psline[linewidth=1.0pt](3,-0.5)(3,1)
\end{pspicture} \
\}.
\end{align}
In this basis, the action of the TL generators on $\StTL{1}[4]$ is
$e_2 \ \psset{xunit=2mm,yunit=2mm}
\begin{pspicture}(0,0)(3,1)
\psline[linewidth=1.0pt](0,-0.5)(0,1)
\psellipticarc[linecolor=black,linewidth=1.0pt]{-}(1.5,1.0)(0.5,1.42){180}{360}
\psline[linewidth=1.0pt](3,-0.5)(3,1)
\end{pspicture} = \fug \ \psset{xunit=2mm,yunit=2mm}
\begin{pspicture}(0,0)(3,1)
\psline[linewidth=1.0pt](0,-0.5)(0,1)
\psellipticarc[linecolor=black,linewidth=1.0pt]{-}(1.5,1.0)(0.5,1.42){180}{360}
\psline[linewidth=1.0pt](3,-0.5)(3,1)
\end{pspicture}\ $, 
$e_2 \ \psset{xunit=2mm,yunit=2mm}
\begin{pspicture}(0,0)(3,1)
 \psellipticarc[linecolor=black,linewidth=1.0pt]{-}(0.5,1.0)(0.5,1.42){180}{360}
\psline[linewidth=1.0pt](2,-0.5)(2,1)
\psline[linewidth=1.0pt](3,-0.5)(3,1)
\end{pspicture} = \ \psset{xunit=2mm,yunit=2mm}
\begin{pspicture}(0,0)(3,1)
\psline[linewidth=1.0pt](0,-0.5)(0,1)
\psellipticarc[linecolor=black,linewidth=1.0pt]{-}(1.5,1.0)(0.5,1.42){180}{360}
\psline[linewidth=1.0pt](3,-0.5)(3,1)
\end{pspicture} \ $, and $e_3 \ \psset{xunit=2mm,yunit=2mm}
\begin{pspicture}(0,0)(3,1)
 \psellipticarc[linecolor=black,linewidth=1.0pt]{-}(0.5,1.0)(0.5,1.42){180}{360}
\psline[linewidth=1.0pt](2,-0.5)(2,1)
\psline[linewidth=1.0pt](3,-0.5)(3,1)
\end{pspicture} = 0$. To give a complete example, in the basis $\StTL{0}[4] =
\{ \
\psset{xunit=2mm,yunit=2mm}
\begin{pspicture}(0,0)(3,1)
 \psellipticarc[linecolor=black,linewidth=1.0pt]{-}(1.5,1.0)(1.5,1.42){180}{360}
 \psellipticarc[linecolor=black,linewidth=1.0pt]{-}(1.5,1.0)(0.5,0.71){180}{360}
\end{pspicture}
,\, \psset{xunit=2mm,yunit=2mm}
\begin{pspicture}(0,0)(3,1)
 \psellipticarc[linecolor=black,linewidth=1.0pt]{-}(0.5,1.0)(0.5,1.42){180}{360}
 \psellipticarc[linecolor=black,linewidth=1.0pt]{-}(2.5,1.0)(0.5,1.42){180}{360}
\end{pspicture} \
\}$, the full action of the TL generators is given by
\begin{equation}
e_1 = \left(\begin{array}{cc} 0 & 0  \\ 1 & n  \end{array} \right), \ \ e_2 = \left(\begin{array}{cc} \fug & 1  \\ 0 & 0  \end{array} \right), \ \ e_3 = \left(\begin{array}{cc} 0 & 0  \\ 1 & n  \end{array} \right).
\end{equation}  

When $\q={\rm e}^{i\pi/p}$ is a root of unity, the situation becomes much more complicated.
The first striking feature is that the standard modules become reducible, but indecomposable -- that is,
there is no way to decompose them onto irreducible representations. As an example, let us consider
the standard module $\StTLn{0}{4}$ with basis
$\StTLn{0}{4}=\{\psset{xunit=2mm,yunit=2mm}
\begin{pspicture}(0,0)(3,1)
 \psellipticarc[linecolor=black,linewidth=1.0pt]{-}(0.5,1.0)(0.5,1.42){180}{360}
 \psellipticarc[linecolor=black,linewidth=1.0pt]{-}(2.5,1.0)(0.5,1.42){180}{360}
\end{pspicture},\begin{pspicture}(0,0)(3,1)
 \psellipticarc[linecolor=black,linewidth=1.0pt]{-}(1.5,1.0)(1.5,1.42){180}{360}
 \psellipticarc[linecolor=black,linewidth=1.0pt]{-}(1.5,1.0)(0.5,0.71){180}{360}
\end{pspicture}  \}$.
When $\q=\mathrm{e}^{i \pi/3}$ ($\fug = 1$), it is easy to see that the space $\IrrTL{2}  =\{\psset{xunit=2mm,yunit=2mm} \begin{pspicture}(0,0)(3,1)
 \psellipticarc[linecolor=black,linewidth=1.0pt]{-}(0.5,1.0)(0.5,1.42){180}{360}
 \psellipticarc[linecolor=black,linewidth=1.0pt]{-}(2.5,1.0)(0.5,1.42){180}{360}
\end{pspicture} - \begin{pspicture}(0,0)(3,1)
 \psellipticarc[linecolor=black,linewidth=1.0pt]{-}(1.5,1.0)(1.5,1.42){180}{360}
 \psellipticarc[linecolor=black,linewidth=1.0pt]{-}(1.5,1.0)(0.5,0.71){180}{360}
\end{pspicture}  \}$ is invariant under the action of $\TLq{4}$. The module 
 $\StTLn{0}{4}$ is thus reducible but indecomposable, and we represent its structure
 by the following diagram
\begin{align}
\StTLn{0}{4} &= \quad \IrrTL{0} \longrightarrow \IrrTL{2}, \notag \\
             &= \quad \{ \psset{xunit=2mm,yunit=2mm}\begin{pspicture}(0,0)(3,1)
 \psellipticarc[linecolor=black,linewidth=1.0pt]{-}(0.5,1.0)(0.5,1.42){180}{360}
 \psellipticarc[linecolor=black,linewidth=1.0pt]{-}(2.5,1.0)(0.5,1.42){180}{360}
\end{pspicture}  \} \longrightarrow \{ \psset{xunit=2mm,yunit=2mm}\begin{pspicture}(0,0)(3,1)
 \psellipticarc[linecolor=black,linewidth=1.0pt]{-}(0.5,1.0)(0.5,1.42){180}{360}
 \psellipticarc[linecolor=black,linewidth=1.0pt]{-}(2.5,1.0)(0.5,1.42){180}{360}
\end{pspicture} - \begin{pspicture}(0,0)(3,1)
 \psellipticarc[linecolor=black,linewidth=1.0pt]{-}(1.5,1.0)(1.5,1.42){180}{360}
 \psellipticarc[linecolor=black,linewidth=1.0pt]{-}(1.5,1.0)(0.5,0.71){180}{360}
\end{pspicture}  \} .
\end{align}
The arrow in these diagrams (``subquotient structure'') should be understood as the action of TL on $\StTLn{0}{4}$. It means 
that it is possible to go from $\{ \psset{xunit=2mm,yunit=2mm}\begin{pspicture}(0,0)(3,1)
 \psellipticarc[linecolor=black,linewidth=1.0pt]{-}(0.5,1.0)(0.5,1.42){180}{360}
 \psellipticarc[linecolor=black,linewidth=1.0pt]{-}(2.5,1.0)(0.5,1.42){180}{360}
\end{pspicture}  \}$ to $\IrrTL{2}$ acting with TL generators, but not the other
way around. To be more precise, it means that $\IrrTL{2}$ is an irreducible submodule
in $\StTLn{0}{4}$, and the quotient $\StTLn{0}{4}/\IrrTL{2} \cong \IrrTL{0} $ by
this submodule is also irreducible.

This structure is quite general, and it can be shown that other standard modules
have a similar indecomposable pattern for other roots of unity. These results
can be found in~\cite{M1,M0,Westbury,MWood} (see~\cite{GV} for complete results using techniques 
similar to those developed in this paper and see also a recent paper~\cite{RidS}). We will only give the main results here 
and refer the reader to those references for details and proofs. It turns out that the irreducible
(also called simple) modules $\IrrTL{j}$ of the Temperley-Lieb algebra when $\q=\mathrm{e}^{i \pi/p}$ is a root of unity
can still be labeled by $0\leq j\leq N/2$.
The standard modules can then be indecomposable, with the following subquotient structure 
\begin{equation}\label{TL-St-struct}
\StTL{j}:\quad\IrrTL{j}\longrightarrow\tIrrTL{j+p-1-2(j\modd p)}
\qquad \text{where}\quad\tIrrTL{j'}=
\begin{cases}
\IrrTL{j'},\quad&\text{if}\;j'>j,\\
0,\quad&\text{if}\;j'=j,\\
\IrrTL{j'+p},\quad&\text{if}\;j'<j,
\end{cases}
\end{equation}
where $0\leq(j\modd p)<p$ is the remainder
and we additionally set $\StTL{j}=0$ for all $j>N/2$ which is crucial
when the number of through lines $2j$ is close to its maximum value
$2j=N$. We note also that the standard modules are irreducible whenever
$j\modd p = \frac{kp-1}{2}$ with $k=1,2$. In particular, all the
standard modules are irreducible for $p=2$ and odd $N$, {\it i.e}, for all half-integer values of $j$. The subquotient structure~\eqref{TL-St-struct} then allows to compute
the dimension $\dd^0_j$ of the irreducible modules taking standard
alternating sums:
\begin{equation}\label{dimIrr}
\dim(\IrrTL{j}) \equiv \dd^0_j = \sum_{n\geq0}\dd_{j+np} - 
\sum_{n\geq t(j)+1}\dd_{j+np-1-2(j\modd p)}.
\end{equation}
where we recall that $\dd_{j}$ is given by~\eqref{eqDimStdTL} and
 we also introduce the step function $\stf(j)\equiv\stf$ as
\begin{equation}
\stf=
\begin{cases}\label{stf-def}
1,&\text{for}\quad  j\modd p> \ffrac{p-1}{2},\\
0,&\text{for}\quad  j\modd p< \ffrac{p-1}{2}.
\end{cases}
\end{equation}

\medskip

We then introduce more complicated modules which are gluings of  a pair of the standard modules $\StTL{j}[N]$
just described. These will be ``fundamental blocks'' or indecomposable direct summands in the XXZ spin chains
and they are technically called \textit{tilting} modules.

 All tilting modules should be filtered by (or composed of) the standard modules and satisfy a self-duality condition~\cite{Donkin}, {\it i.e.}, they should be invariant under the adjoint $\cdot^{\dagger}$ operation  (see also~\cite{GJSV} for a short review in the  context of boundary spin chains.) 
 What is important for us is that the structure of tilting modules can be in principle deduced from these conditions~\cite{M0}  though in general it is a very hard problem\footnote{For the present case of TL algebras, this problem is not too hard. To construct the tilting modules one should know first-extension groups between different standard modules. To find them it is enough to describe the filtration of projective covers by the standard modules. This part can be easily done using the quasi-hereditary structure of the TL algebra and the reciprocity property for projective covers, at least for the cases $p\ne2$. This step was also explained in~\cite{GV}. Having the extension groups we can study the tilting modules step by step: (i) extend a (reducible) standard module $\StTL{j}$ by another standard $\StTL{j'}$ with the  minimum value of $|j-j'|$ (ii) then do the same with $\StTL{j'}$. The result at some step might become a decomposable module. The self-duality and uniqueness properties of the tilting modules actually stop the construction at the first step because we obtain projective covers.}.

It is known from the general theory~\cite{Donkin} of tilting modules that
 there is a one-to-one correspondence with the standard modules.
 We thus introduce  $\PrTL{j}$ as {\sl the indecomposable} tilting module that can be mapped onto the standard module $\StTL{j}$.
First, those tilting modules corresponding to $j\modd p = \frac{kp-1}{2}$ with $k=1,2$ are the standard modules and are irreducible.
For other cases when the standard modules are not irreducible,
we use the diagram for their subquotient structure~\eqref{TL-St-struct} and see which ones can be glued in order to produce a self-dual module, {\it i.e.}, a  tilting module. As we found such a module we are done because of the uniqueness. In the case of TL algebras, the construction ends at the first step and we obtain a gluing of only two standard modules. We thus have,
for $j$ integer or half integer, the subquotient structure for the indecomposable tilting modules in terms of standards:
\begin{align}
&\PrTL{j}:\quad\StTL{j}\longrightarrow\tStTL{j-1-2(j\modd p)},& {\rm for} \ \ \ffrac{p}{2}\leq j,&
\end{align}
where we set $j\modd p\ne \frac{kp-1}{2}$ with $k=1,2$ and
\begin{equation}\label{TL-tSt-def}
\tStTL{j'}=
\begin{cases}
\StTL{j'},\quad&\text{if}\;j'+p>j,\\
0,\quad&\text{if}\;j'+p=j,\\
\StTL{j'+p},\quad&\text{if}\;j'+p<j.
\end{cases}
\end{equation}
In terms of simple modules, we finally get the structure of subquotients, for
 $j \geq p/2$, as
\begin{align}\label{prTL-pic-gen-dense-even}     
   \xymatrix@C=5pt@R=15pt@M=2pt{%
    &&\\
    &\PrTL{j}: &\\
    &&
 }      
&  \xymatrix@C=2pt@R=15pt@M=2pt@W=2pt{%
    &&{\IrrTL{j}}\ar[dl]\ar[dr]&\\
    &\IrrTL{j+\stf p-1-2(j\modd p)}\ar[dr]&&\IrrTL{j+(1+\stf)p-1-2(j\modd p)}\ar[dl]\\
    &&\IrrTL{j}&
 } \quad
&   \xymatrix@C=5pt@R=15pt@M=2pt{%
    &&\\
    &&\\
    &&
 }&      
\end{align}
where the most right subquotient is absent whenever its subscript $j+(1+\stf)p-1-2(j\modd p)$ is greater than $N/2$; so, the tilting module has only three subquotients in this case. We refer to~\cite{GV} for more details. 
Explicit examples of such indecomposable modules will arise naturally when we turn to the computation of lattice fusion rules (see section~\ref{secFusion}).

These tilting modules $\PrTL{j}$ with the ``diamond''-type diagram in~\eqref{prTL-pic-gen-dense-even} are actually the only TL modules
we need in order to describe fully the structure of our spin-chains as bimodules.
We then review these bimodules  but first shortly discuss the general idea.

\subsection{Spin chain structure at roots of unity}
The representation theory of the symmetry algebra $\cent_A$ is usually much
easier to study than  the representation theory of the  ``hamiltonian
densities" algebra $A$. It is thus more reasonable to start with a
decomposition of spin-chains over $\cent_A$ into indecomposable direct summands, which are again tilting modules~\cite{Donkin}. 
The next step is to  study all
homomorphisms or intertwining operators between the direct summands in the decomposition 
 to obtain  the module structure over the ``hamiltonian densities''
algebra $A$. In particular, multiplicities in front of tilting
$\cent_A$-modules give the  dimensions of simple $A$-modules, and the
subquotient structure of tilting $A$-modules can be deduced from
the one of the tilting $\cent_A$-modules, see~\cite{AF-book}.
As a result, one  gets a sequence of bimodules $\Hilb_N$ over the two commuting algebras
parametrized by the number $N$ of sites/tensorands in the
spin-chain. 

\newcommand{\XX}{\mathcal{X}}
Following these lines, we
obtain finally the decomposition of the spin-chain $\Hilb_{N}$ over $\TLq{N}$ as~\cite{GV}
\begin{multline}\label{decomp-TL}
\Hilb_{N}|_{\rule{0pt}{7.5pt}%
\TLq{N}} \cong \bigoplus_{r=1}^{r_m-1} \bigoplus_{\substack{s=0,\\rp+s+N=1\modd 2}}^{p-1} \dim\bigl(\XX_{p-s,r}\bigr)\PrTL{\frac{rp+s-1}{2}}
\oplus\bigoplus_{\substack{s=0,\\s+s_m=1\modd 2}}^{s_m+1} \dim\bigl(\XX_{p-s,r_m}\bigr)\PrTL{\frac{r_m p+s-1}{2}}\\
\oplus\bigoplus_{\substack{s=1,\\s+s_m=1\modd 2}}^{s_m+1} \dim\bigl(\XX_{s,r_m+1}\bigr)\IrrTL{\frac{r_m p+s-1}{2}}
\oplus\bigoplus_{\substack{s=s_m+2,\\s+s_m=1\modd 2}}^{p-1} \dim\bigl(\XX_{p-s,r_m}\bigr)\IrrTL{\frac{r_m p-s-1}{2}},
\end{multline}
where we recall that $N=r_m p+s_m$, for $r_m\in\oN$ and $-1\leq s_m\leq p-2$. Here, we use the notation $\XX_{s,r}$ for irreducible representations  of the quantum group $\LQG$. They have dimension $rs$ and they are irreducible quotients of the spin-$n$ quantum group representations, where $n=\frac{p(r-1)+s-1}{2}$. We will also use the notation $\XX_{[n]}$.

\begin{figure}
\begin{center}
\includegraphics[scale=1.2]{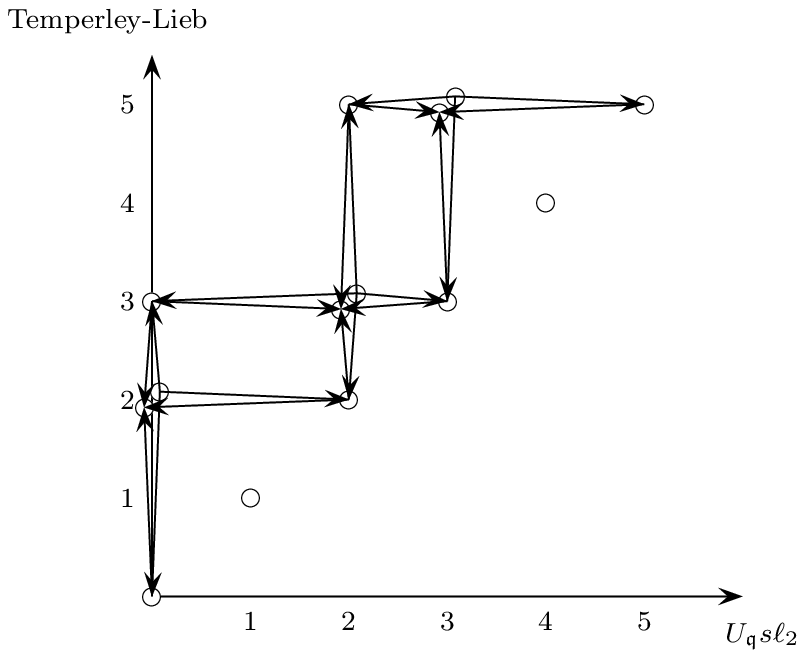}
\end{center}
  \caption{Bimodule for percolation ($\q=\mathrm{e}^{i \pi/3}$) and $N=10$ sites. Horizontal ({\it resp.} vertical) arrows correspond to the action of the quantum group $\LQG$ ({\it resp.} the Temperley-Lieb algebra). Each node with a Cartesian coordinate $(n,n')$ corresponds to the tensor product $\IrrTL{n'}\otimes\XX_{[n]}$. Some nodes
   occur twice and those nodes
  have been separated slightly for clarity.}
  \label{figStaircasePerco}
\end{figure}

Just like in semisimple cases, it is convenient to represent the Hilbert space structure as a bimodule over both Temperley-Lieb and $\LQG$~\cite{RS3}.
As an example, we show in Fig.~\ref{figStaircasePerco} the analogue of Fig.~\ref{aba:fig4} for $\q=\mathrm{e}^{i \pi/3}$ ($\fug=1$, percolation) at $N=10$ sites. The decomposition over the TL algebra is
\begin{equation}
\Hilb_{10}|_{\rule{0pt}{7.5pt}%
\TLq{10}} \cong 3 \PrTL{1}
\oplus \PrTL{2}\oplus 4\PrTL{3} \oplus 9\PrTL{4} \oplus 3\PrTL{5}\oplus 8\IrrTL{5}.
\end{equation}
 In  the bimodule diagram, each node with a Cartesian coordinate $(n,n')$ 
corresponds to the tensor product $\IrrTL{n'}\otimes\XX_{[n]}$  of simple modules over the TL algebra and $\LQG$, respectively, and
the arrows show the action of both algebras -- the Temperley--Lieb $\TLq{N}$ acts
in the vertical direction (preserving the coordinate $n$), while
$\LQG$ acts in the horizontal direction. The diamond-shape tilting $\TLq{N}$-modules
$\PrTL{n'}$ described in~\eqref{prTL-pic-gen-dense-even}  can be recovered by ignoring all the
horizontal arrows
of the bimodule diagram. These are squeezed, so that the first tilting TL module $\PrTL{1}$ is just a node (this one is irreducible), the second tilting $\PrTL{2}$ consists of the left-most set of four vertical arrows, etc.

\section{General strategy: scaling limit and bimodules}\label{sec:gen-lim}

The idea is now to analyze the spin chain from an algebraic point of view, with the motivation that the algebra of local energy hamiltonian densities should go over, in the continuum limit to the Virasoro algebra, and that many of its features may be stable as the length of the chain is increased, as long as one focuses only on low energy excitations.
So our general strategy will be to consider the XXZ spin chain~\eqref{XXZ_H} as a lattice regularization for (L)CFTs.
The representation theory of the TL algebra when $\q$ is a root of unity then mimics what happens in the scaling
limit for the Virasoro algebra. One can even obtain interesting results for the Virasoro algebra representation
theory, starting directly from lattice models. The idea of doing so probably goes back
to \cite{PasquierSaleur}, and was pushed forward recently by Read and Saleur, who studied the structure of XXZ spin chains
and supersymmetric models~\cite{RS2,ReadSaleur01} on the lattice.

\subsection{Scaling limit and Virasoro algebra}

It is not  clear how the continuum limit can be taken in a
mathematically rigorous way for any~$\q$, but roughly speaking, we take the
eigenvectors of $H$ in the spin-chain that have low-energy eigenvalues only, and we
expect that the inner products among these vectors can be made to tend
to some limits. Further, if we focus on long wavelength Fourier
components of the set of local generators $e_j$, we expect their limits to exist, and their commutation
relations to tend to those of the Virasoro generators $L_n$ (this was shown explicitly for free fermion systems: for the Ising chain in~\cite{KooSaleur}, and  for the XX model in~\cite{GRS1}), in the sense of strong
convergence of operators in this basis of low-energy
eigenvectors. Then, the modules over the TL algebra restricted to
the low-energy states become, now in the scaling limit\footnote{The two notions -- continuum and scaling limits -- are essentially the same and below we will mostly use the more algebraic one which is the scaling limit; see also a similar discussion in~\cite{GRS1}.}, modules over the Virasoro algebra at appropriate central charge.

As an example, let us discuss how the TL standard modules become  so-called Kac modules over the Virasoro algebra when the
scaling limit is taken~\cite{PasquierSaleur}. For $\q=\mathrm{e}^{i \pi/p}$ ($p \in \mathbb{R}$ here), we introduce the following formula
for the central charge
\begin{equation}\label{eqCentralCharge}
\displaystyle c_{p-1,p} = 1 - \frac{6}{p (p-1)}.
\end{equation}
The Kac formula at central charge $c_{p-1,p}$ reads
\begin{equation}
\displaystyle h_{r,s} = \frac{ \left(p r - (p-1)s \right)^2 - 1}{4 p (p-1)}.
\end{equation}
Using Bethe ansatz and keeping only low-lying excitations, it can be then shown that the  spectrum generating function of the module $\StTL{j}[N]$ has the following limit~\cite{PasquierSaleur}
\begin{equation}\label{gen-func-lim}
\displaystyle \lim_{N\to\infty} \sum_{{\rm states} \ i} q^{\frac{N}{\pi v_F} \left(E_i(N) - N e_{\infty} \right)} = q^{-c/24} \  \dfrac{q^{h_{1,1+2j}}-q^{h_{1,-1-2j}}}{\prod_{n=1}^{\infty} \left( 1 - q^n\right)},
\end{equation}
where $v_F=\frac{\pi \sin \gamma}{\gamma}$ is the Fermi velocity, $2\cos{\gamma}=\q+\q^{-1}=\fug$ is the fugacity of a loop, $c=c_{p-1,p}$ is the central charge, $E_i(N)$ is the eigenvalue of the $i^{th}$ (counted from the vacuum) eigenstate of $H=-\sum_i e_i$, and $e_{\infty} = \lim_{N \rightarrow \infty} E_0(N)/N$, with $ E_0(N)$ the groundstate energy. The expression on the right-hand side of~\eqref{gen-func-lim} coincides 
with the Virasoro character ${\rm Tr}\, q^{L_0 - c/24}$ of the \textit{Kac module} with conformal weight $h_{1,1+2j}$ defined
as a quotient of the covering Verma module as $\VK_{1,1+2j} \equiv \Verma_{h_{1,1+2j}}/\Verma_{h_{1,-1-2j}}$.
We use here the standard notation $\Verma_{h}$ for the Virasoro Verma module generated from the highest-weight state of weight $h$~\cite{YellowBook}.
We already see at the level of generating functions and characters that we have a deep correspondence between the TL  and Virasoro algebras in the scaling limit,
where the (properly rescaled) Hamiltonian $H$ becomes the $L_0$ generator. As mentioned above, it is even possible to construct
other Fourier modes by taking appropriate combinations of TL generators on the lattice that will tend (in a sense that can be made rigorous in some cases) to other Virasoro generators $L_n$ in the limit~\cite{KooSaleur,GRS1}. Thanks to different techniques (numerical or analytical whenever possible) it can be shown that the lattice operators
\begin{equation}
\displaystyle L^{(N)}_n = \frac{N}{\pi} \left[ - \frac{1}{v_F} \sum_{k=1}^{N-1} (e_k - e_{\infty}) \cos \left( \frac{n k \pi}{N} \right) +  \frac{1}{v_F^2}  \sum_{k=1}^{N-2} \left[ e_k, e_{k+1} \right] \sin \left( \frac{n k \pi}{N} \right)\right] + \frac{c}{24} \delta_{n,0},
\label{eq_KooSaleur}
\end{equation}
become the Virasoro modes $L_n$ in the continuum limit at $N\to \infty$. We emphasize that although this formula is a conjecture -- its derivation in~\cite{KooSaleur} is definitely heuristic, its validity is quite well established. It is indeed possible to show that this expression converges to the usual Virasoro generators in the case of non-interacting fermionic systems~\cite{GRS1,GRS3,GST}, and there is now a fair amount of numerical results~\cite{KooSaleur,DJS,VJS,VGJS} on Virasoro inner products in interacting systems (including the measure of indecomposability parameters to be discussed below) that tend to show that this formula remains valid even in the presence of interactions.

\subsection{Hilbert space structure and bimodules in the limit}

\begin{figure}
\begin{center}
\includegraphics[scale=1.3]{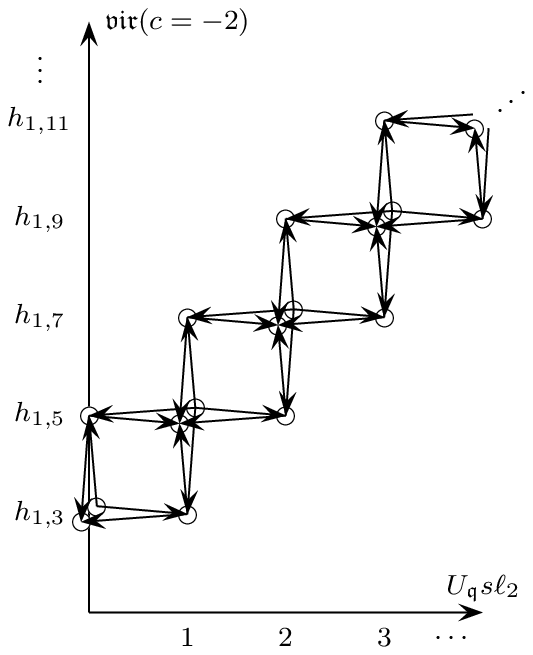}
\hspace*{1cm}
\includegraphics[scale=1.3]{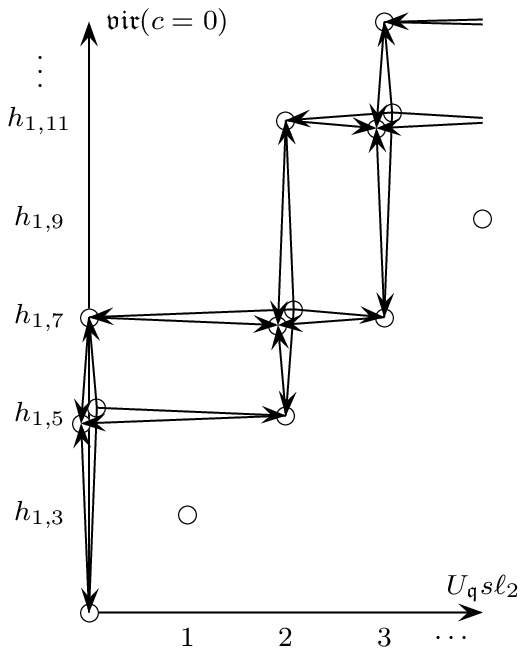}
\end{center}
  \caption{Bimodules for dense polymers ($p=2$, $\q=i$ and $c=-2$) and boundary percolation ($p=3$, $\q=\mathrm{e}^{i\pi/3}$ and $c=0$) showing the commuting action of the Virasoro algebra and the quantum group $\LQG$ (see~\cite{RS3}). The labels along the horizontal axis correspond to the $\LQG$ spin $j$ (also called $n$ previously). }
  \label{openbimodule-cont}
\end{figure}

Having now the structure of the spin-chain as a (bi)module over the two commuting algebras for each finite $N$, we can analyze its behavior in the limit $N\to\infty$.
It is clear that the symmetry algebra of the Hamiltonian densities also provides a symmetry of the low-lying part of spectrum of the Hamiltonian.
The symmetry  algebra in the scaling limit, which commutes now with the Virasoro algebra, must be thus at least as large as that in the finite-$N$ chains. The only difference in the limit is that we now admit  arbitrarily high values of the $\LQG$ spins.
For example,
the decomposition of the open XXZ spin-chain as a bimodule over the pair
$(\TLq{N},\LQG)$ of commuting algebras, like in Fig.~\ref{figStaircasePerco}, goes over in the
scaling limit to a semi-infinite (`staircase') bimodule over the
Virasoro algebra $\Vir(c_{p-1,p})$, with central charge $c_{p-1,p}$, and an infinite-dimensional representation
of $\LQG$. 
This is illustrated in
Fig.~\ref{openbimodule-cont} for the example of percolation, where the same comments as in the finite
chain apply exactly, with the replacement of $\TLq{N}$ by the Virasoro
algebra. Using the correspondence between the irreducible TL modules $\IrrTL{j}$ and  irreducible Virasoro modules with weight $h_{1,2j+1}$, which holds at least at the level of characters, we obtain complicated indecomposable Virasoro modules that we describe in the next section.


While the scenario described above cannot be analytically established for general models, it is confirmed a posteriori by the
validity of the results (structure of Virasoro modules and their fusion) obtained using the bimodule structure~\cite{RS3,GV}. Of course,
in some special cases, such as free theories, much more can be said. For instance, the associated symplectic fermions CFT arising in the
scaling limit of the XXZ spin-chains at the free fermion point ($n=0$ or $\q=i$) can be analyzed independently of the lattice results. Recall that the symplectic fermions theory action involves two fermionic fields of dimension $0$, and has Noether's currents generating a global $SU(2)$ symmetry~\cite{Kausch}. Together with the fermionic zero modes, we obtain the full symmetry algebra of operators commuting with the Virasoro algebra. It turns out that this symmetry algebra is realized by a representation of the quantum group $\LQG$ at $\q=i$, see~\cite{GRS1}. The full Hilbert space in such chiral LCFT can then be decomposed onto indecomposable Virasoro modules and its symmetry algebra, with precisely the same result as in the $p=2$ analog of Fig.~\ref{openbimodule-cont} (see~\cite{RS3}). It is even possible to show~\cite{GRS1,GST} that the lattice regularizations $L_n^{(N)}$ of the Virasoro modes indeed converge to the well-known~\cite{Kausch} symplectic fermions representation of $L_n$'s.

\section{Indecomposable Virasoro representations from lattice models}

A great deal of progress in  our understanding of LCFTs has come from the abstract study of indecomposable (but not irreducible) modules of the Virasoro algebra. Following the pioneering work of Rohsiepe~\cite{Rohsiepe}, various attempts were made to try to build and classify these modules~\cite{KytolaRidout}, and study their fusion~\cite{GK1,MathieuRidout,GabRunW2,JR}, often running into considerable technical difficulties. As we shall see, the lattice approach turns out to be an extremely efficient way to obtain all this algebraic information, and more. In this section, we review the ideas of~\cite{RS2,RS3,GJSV}, and show how to obtain Virasoro indecomposable modules starting from finite-dimensional representations of lattice algebras. We discuss Virasoro staggered modules and their associated indecomposability parameters from a lattice perspective and discuss how lattice models can be used further to obtain more complicated indecompable modules.

\subsection{Virasoro staggered modules from the lattice}
As discussed in Sec.~\ref{sec:gen-lim}, the continuum limit of the XXZ spin chain at $\q=\mathrm{e}^{i\pi/p}$ is
described by a CFT with central charge $c_{p-1,p}$ given by~\eqref{eqCentralCharge}.
In particular, the generating functions of energy levels on the standard modules $\StTL{j}$
of the TL algebra at $\q=\rme^{i\pi/p}$ give in the limit the characters of
the Kac modules $\VK_{1,2j+1}$ over the Virasoro 
algebra $\Vir(c_{p-1,p})$. Note also that the finite alternating sum~\eqref{dimIrr} for the dimension of the
irreducible TL module $\IrrTL{j}$ corresponds in the limit to an infinite alternating sum
of the Kac characters, giving rise to the well-known Rocha--Caridi formula for the irreducible characters~\cite{RC-paper}.

Furthermore, using our semi-infinite bimodules (see Fig.~\ref{openbimodule-cont} in the example of percolation), we can extract Virasoro modules
keeping only the vertical arrows. 
We then obtain the following diamond-shape diagram for indecomposable Virasoro modules, for $j\modd p\ne \frac{kp-1}{2}$ with
$k=1,2$,
\begin{align}\label{stagg-pic-gen-dense-even}     
   \xymatrix@C=5pt@R=22pt@M=2pt{%
    &&\\
    &\VP_{1,2j+1}: &\\
    &&
 }      
&  \xymatrix@C=4pt@R=22pt@M=3pt@W=3pt{%
    &&{h_{1,2j+1}}\ar[dl]\ar[dr]&\\
    &{\;\;h_{1,2j'-1}\;\;}\ar[dr]&&{\,h_{1,2(j'+p)-1}\,}\ar[dl]\\
    &&{h_{1,2j+1}}&
 } \quad
&   \xymatrix@C=3pt@R=22pt@M=2pt{%
    &&\\
    &\text{for}\quad  j \geq \ffrac{p}{2},&\\
    &&
 }&      
\end{align}
where $j'=(j+\stf(j) p)-2(j\modd p)$ and the function $\stf(j)$ was defined in~\eqref{stf-def}, and we  denote the irreducible Virasoro subquotients simply by their conformal weights $h_{1,j}$. 
 We note that a
south-east arrow represents an action of negative Virasoro modes while
a south-west arrow represents positive modes action. In the
diagram~\eqref{stagg-pic-gen-dense-even}, the indecomposable Virasoro module is a
`gluing'/extension of two indecomposable Kac modules which are
highest-weight modules. The one in the top composed of
irreducibles of the weights $h_{1,2j+1}$ and $h_{1,2(j'+p)-1}$ is the quotient $\Verma_{h_{1,2j+1}}/\Verma_{h_{1,-2j-1}}$ of the Verma module
with the weight $h_{1,2j+1}$ by the singular vector\footnote{Recall that  the Verma module $\Verma_{h_{1,2j+1}}$ with $2j$ integer is reducible with a proper submodule isomorphic to $\Verma_{h_{1,-2j-1}}$.}  at the level $2j+1$,
and the second Kac module in the bottom  is a similar quotient  of the Verma module with
the weight $h_{1,2j'-1}$.

We emphasize that the modules with the structure~\eqref{stagg-pic-gen-dense-even}  obtained using the lattice algebraic analysis indeed exist. They are known under the name \textit{staggered Virasoro modules}. In general, a staggered module is a gluing (an extension) of two highest-weight Virasoro modules with a non-diagonalizable action of $L_0$. 
 A complete theory of staggered modules was developed by Kyt\"{o}l\"{a} and Ridout~\cite{KytolaRidout},
following the pioneering work of Rohsiepe~\cite{Rohsiepe}.

 We see that the
 staggered Virasoro modules for different central charges abstractly discussed in~\cite{Rohsiepe,GK1} and~\cite{KytolaRidout}
 can quickly be recovered from the lattice -- at least their subquotient
 structure can be deduced from our bimodule. Adding to it the conjectured Koo--Saleur formula~\eqref{eq_KooSaleur} for the Virasoro generators, this opens the way to measuring~\cite{DJS,VJS} indecomposability parameters (also called
 $\beta$ invariants~\cite{KytolaRidout}) characterizing Virasoro-module structure completely. Finally,  we will show in the next section how the idea of studying Virasoro modules by taking scaling limits of the spin-chains can be extended in order to compute fusion
 rules using an induction procedure~\cite{RS2,RS3, GV}. In this section, we continue studying the scaling limit of TL modules by measuring their indecomposability parameters on the lattice and introducing bigger lattice algebras that give even more complicated Virasoro modules.

\subsection{Lattice indecomposability parameters}

\begin{table}
\begin{center}
\begin{tabular}{|c|c|c|c|}
  \hline
  $N$ & $b_{\rm boundary}$ & $ N$ & $b_{\rm bulk}$\\
  \hline
  10 & -0.605858 & 10 &  -4.33296\\
   12 & -0.606403 & 12  & -4.55078  \\
   14 & -0.607775 & 14  & -4.68234  \\
   16  & -0.609226 & 16  & -4.76634 \\ 
  18  & -0.610561 & 18  & -4.82256\\ 
   20 & -0.611738  & 20 & -4.86168\\
   22 & -0.612764  & 22 & -4.88978\\
  \hline
 $\infty$ &   -0.6249 $\pm$ 0.0005  & $\infty$ & -5.00 $\pm$  0.01  \\
  \hline
  \hline
  Exact & -5/8 = 0.625 & Exact & $-5$\\
  \hline
\end{tabular}
\end{center}
\caption{Numerical measure of the $b$ parameter in percolation with open~\cite{DJS,VJS} and periodic~\cite{VGJS} boundary conditions.}
  \label{tab_perco}
\end{table}

Virasoro staggered modules are characterized by universal numbers called logarithmic couplings or indecomposability parameters. Indecomposability parameters are universal, and they are believed to play an important role in physical applications of LCFTs. They can be defined rather abstractly~\cite{KytolaRidout,MathieuRidout} as parameters crucial for characterizing the staggered modules completely, or they can be thought of as universal coefficients that appear in front of logarithmic singularities in correlation functions\footnote{In general however, there are some subtle differences between the `algebraic' indecomposability parameters and the coefficients that appear in correlation functions~\cite{MathieuRidout1,GV}.}  of fields living in such modules.

While the analysis of symmetries of the lattice models provides results about the general structure of  the Virasoro indecomposable modules, getting more detailed information about the action of the Virasoro generators in these modules -- such as the  numerical values of the indecomposability parameters -- is more challenging. 

Although the method is completely general, we will focus here on the celebrated $b$-number that  characterizes the logarithmic structure associated with the stress energy tensor at $c=0$~\cite{Gurarie, Gurarie2,GurarieLudwig2} (see Sec.~\ref{subsecCatastrophe}). Recall that the logarithmic partner $t(z)$ of the stress-energy tensor satisfies
\begin{equation}
\langle t(z)t(0)\rangle = {-2b\ln z+a\over z^4},
\end{equation}
where $a$ is an irrelevant constant and the state $\Ket{t}=\lim_{z\to0}t(z)\Ket{0}$ is normalized such that 
\be 
L_0 \Ket{t} = 2 \Ket{t} + \Ket{T}
\ee
 (and $\Ket{T} = L_{-2} \Ket{0}$) in radial quantization. The parameter $b$ can then be expressed as $b=\Braket{T | t}$. This $b$ parameter has attracted a lot of attention since it was introduced by Gurarie, and computing the values allowed for the parameter $b$ in any given $c=0$ conformal field theory, for example the LCFT describing the transition between plateaus in the IQHE, remains an interesting open problem. From an analytical point of view, indecomposability parameters such as $b$ can be computed using algebraic methods~\cite{KytolaRidout,MathieuRidout}, or using heuristic limit arguments~\cite{VJS}.  

For simple $c=0$ theories, namely Self-Avoiding Walks (SAWs also known as dilute polymers) or percolation, $b$ is now known both in the bulk and at the boundary CFTs~\cite{MathieuRidout,VGJS}. It is interesting to notice that $b$ can be directly measured on the lattice, just like the central charge or the conformal dimensions. For percolation ($\q = \mathrm{e}^{i \pi/3}$) for example, the logarithmic structure for the stress-energy tensor corresponds on the lattice to a Jordan cell involving the state $\Ket{T^{(N)}}$ associated\footnote{$\Ket{T^{(N)}}$ is the only state corresponding to the conformal weight $h=2$ in the vacuum sector.} with $T(z)$ in the spectrum of the Hamiltonian $H = - \sum_{i=1}^{N-1} e_i$.  We normalize the states such that in the basis $(\Ket{T^{(N)}},\Ket{t^{(N)}})$, the Hamiltonian reads
\begin{equation}
\displaystyle H^{(N)} - E_0(N) \one= \frac{\pi v_F}{N} \left( \begin{array}{cc} h^{(N)} & 1  \\ 0 & h^{(N)}  \end{array} \right),
\label{eq_latticeHnormJordan}
\end{equation}
where $ E_0(N)$ is the groundstate energy and $h^{(N)} = \frac{N}{\pi v_F} (E(N) - E_0(N)) $, with $\lim_{N \to \infty} h^{(N)} = 2$. This Jordan cell appears because $H$ is not diagonalizable on the Temperley-Lieb tilting module $\PrTL{2}$ described by~\eqref{prTL-pic-gen-dense-even}  
\begin{align*}   
   \xymatrix@C=5pt@R=18pt@M=2pt{%
    &&\\
    &\PrTL{2}: &\\
    &&
 }      
&  \xymatrix@C=4pt@R=18pt@M=3pt@W=3pt{%
    &&{\IrrTL{2}}\ar[dl]\ar[dr]&\\
    &\IrrTL{0}\ar[dr]&&\IrrTL{3}\ar[dl]\\
    &&\IrrTL{2}&
 } 
   \qquad    
   \xymatrix@C=5pt@R=18pt@M=2pt{%
    &&\\
    & \underset{N \rightarrow \infty}{\longrightarrow}  &\\
    &&
 } 
 \qquad
 \xymatrix@C=6pt@R=18pt@M=3pt@W=3pt{%
   &&{t}\ar[dl]\ar[dr]&\\
    &\one \ar[dr]&& \xi \ar[dl]\\
   && T &
 } \quad
&   \xymatrix@C=5pt@R=15pt@M=2pt{%
    &&\\
    &&\\
    &&
 }&   
\end{align*}
In the scaling limit, this module goes to a Virasoro staggered module where the state $t$ lives at the top and $T = L_{-2} \one$ at the bottom (we loosely denote the Virasoro simple modules by the corresponding field). Note that the field $\xi$ has dimension $h_{1,7}=5$. This staggered module is known to be characterized by a number $b=\Braket{T | t}= - \frac{5}{8}$ from algebraic methods~\cite{MathieuRidout}. It is interesting to check this result directly on the lattice. This was first done using a specific trick in~\cite{DJS}, and generalized to many other cases in~\cite{VJS}. The idea is to compute the inner product $\Braket{T | t}$ on the lattice, the main issue being the proper normalization of $\Ket{T^{(N)}}$ which is non-trivial because $\Braket{T^{(N)}|T^{(N)}} = \Braket{T|T}=0$ exactly. A proper normalization is provided by a regularization of the stress energy tensor given by the lattice versions $L_n^{(N)}$ of the Virasoro modes~\eqref{eq_KooSaleur}.

There are actually two crucial steps to measure indecomposability parameters on the lattice: identifying a lattice inner product that will go to the Virasoro bilinear form in the limit, and, as mentioned above, properly normalizing the null state $\Ket{T}$ on the lattice. The Virasoro form on the lattice can be regularized in terms of the TL inner product (non-definite positive!) defined by $e_i^\dag=e_i$. More precisely, for the XXZ spin chain,  this inner product is the usual bilinear form without complex conjugation, that is, treating $\q$ as a formal parameter. For example, on $N=4$ sites, the vector $\Ket{\phi} = \Ket{\uparrow \uparrow \downarrow \downarrow} + \q \Ket{\uparrow \uparrow \uparrow \uparrow} $ has norm  $\Braket{\phi|\phi} = 1+\q^2$.

The second step is the proper normalization of $\Ket{T^{(N)}}$. This is achieved using the Koo-Saleur formula~\eqref{eq_KooSaleur}. Let us define
\begin{equation}
\displaystyle b^{(N)} = \dfrac{\left| \Braket{t^{(N)} \left|L_{-2}^{(N)} \right| 0^{(N)}} \right|^2 }{\Braket{t^{(N)}|T^{(N)}}},
\label{eq_b_lattice}
\end{equation}
where $\Ket{0^{(N)}}$ is the groundstate of the system, and $L_{-2}^{(N)}$ is given by~\eqref{eq_KooSaleur}. It is easy to see that this quantity does not depend on the normalization of $\Ket{T^{(N)}}$, and that it provides a lattice version of $b$.

The various steps to compute $b$ can thus be summarized in the following way:
\begin{enumerate}
\item Using exact diagonalization methods, find a Jordan basis for the first few excitations of $H$ on $N$ sites, with $N$ even.
\item Identify a Jordan cell in the spectrum of $H$ and normalize the states like in eq.~\eqref{eq_latticeHnormJordan}.
\item Also identify the (ground)state $\Ket{0^{(N)}}$ and normalize it such that $\Braket{0^{(N)}|0^{(N)}}=1$ for the lattice inner product.
\item Using Virasoro generators on the lattice~\eqref{eq_KooSaleur}, construct the operator $L^{(N)}_{-2}$.
\item Compute $b^{(N)}$ using eq.~\eqref{eq_b_lattice}.
\end{enumerate}
The value of the indecomposability parameter $b = \lim_{N \rightarrow \infty} b^{(N)}$ is then computed using an extrapolation $b^{(N)} = b + A/N + B/N^2 + \dots$ Numerical results are given in Tab.~\ref{tab_perco}, and are in good agreement with the expectation $b= - \frac{5}{8}$. This approach can in principle be generalized to measure any indecomposability parameter~\cite{VJS}.

More interestingly, the same numerical measurement was done in the bulk~\cite{VGJS} (corresponding now to periodic boundary conditions on the lattice, see below), with a rather unexpected result $b=-5$ (see Tab.~\ref{tab_perco}), in contradiction with earlier expectations~\cite{GurarieLudwig2}. In that case, the numerical result came before a correct theoretical prediction, although this bulk value $b=-5$ is now explained using both heuristic limit arguments~\cite{VGJS} and Virasoro algebra representation theory~\cite{RidoutBulk} -- note that there are two copies of the Virasoro algebra in the bulk, and this usually leads to more complicated indecomposable structures than in the boundary case.

\subsection{Towards a classification of Virasoro indecomposable modules: Blob algebra}

\begin{figure}
\begin{center}
\psfig{file=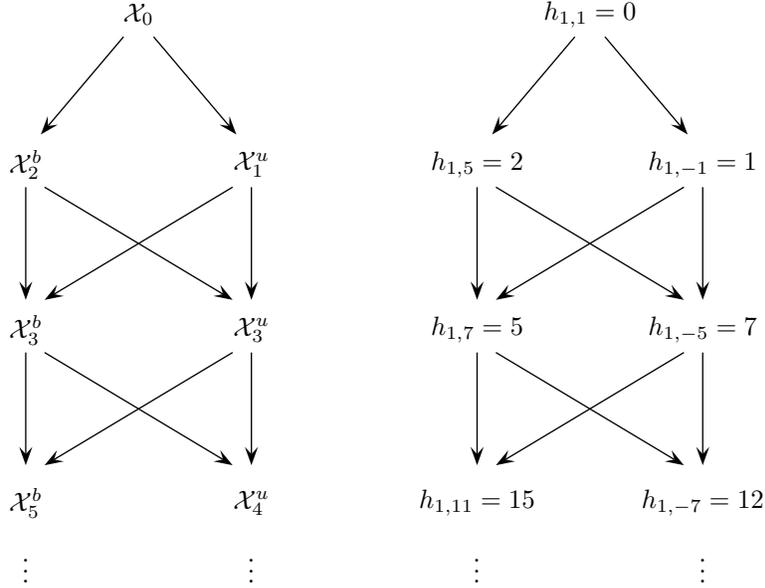}
\end{center}
\caption{Example of standard module $\BW_{0}$ for the blob algebra with $\fug=1$ and $y=1$, and corresponding $c=0$ Verma module in the scaling limit.}
\label{aba:figVerma}
\end{figure}

Let us also briefly mention that the Temperley-Lieb algebra is not the end of the story. Of course, there are many other lattice algebras that one can use to construct statistical models described by Conformal Field Theory. An especially interesting example is provided by the so-called \textit{blob algebra}~\cite{MartinSaleur} (also known as ``one-boundary TL algebra''), as it bears some striking resemblances with the Virasoro algebra from the point of view of its representation theory~\cite{GJSV}.
To define the blob algebra $\mathcal{B}(N,n,y)$, let us start from the Temperley-Lieb algebra and consider  all the words 
written with the $N-1$ generators $e_i$ and an extra ``blob'' generator $b$, subject to the additional relations
\begin{subequations} \label{Blobdef}
\begin{eqnarray}
b^2 &=& b,\\
e_1 b e_1 &=& y e_1,\quad y\in\mathbb{R}, \\
\mbox{}[b,e_i] &=& 0,\qquad i>1.
\end{eqnarray}
\end{subequations}
The extra boundary 
operator $b$ can be interpreted as  decorating strands at the left boundary with a ``blob''. It gives to the corresponding ``blobbed loops''  a weight $y$~\cite{MartinSaleur}, different from the bulk weight $\fug$.

As for the Temperley-Lieb algebra, one can define standard modules $\BW^{b}_{j}$ and $\BW^{u}_{j}$ that are still parametrized by the number of through-lines $2 j$, but there are also two sectors {\sl blobbed} and {\sl unblobbed} corresponding to the two projectors $b$ and $1-b$, respectively~\cite{JSCombBlob}. It is possible to define critical lattice models based on this blob algebra, and one finds that whereas the TL standard modules were related in the scaling limit to Kac modules over the Virasoro algebra, the blob standard modules tend to Verma modules~\cite{JSBlob, GJSV}. The blob algebra is clearly larger than TL and so are its representations, since they correspond in the limit to Verma modules without any quotient being taken. As a consequence, the blob algebra representation theory is much richer~\cite{MartinSaleur, Martin}, and standard modules have a complicated indecomposable structure in non-generic cases (see Fig.~\ref{aba:figVerma} for an example taken from~\cite{GJSV}, where $\BX^{b/u}_{j}$ are simple modules of the blob algebra). The important point is that the blob algebra somehow provides a lattice version of the Virasoro algebra. Though the lattice expressions $L_n^{(N)}$ for $L_n$'s were  proposed~\cite{GJSV} for this bigger algebra as well, this sentence should be understood as a conjecture, since the correspondence has been established only at the level of modules so far. Nevertheless, we can use this correspondence to obtain new results~\cite{GJSV} for the Virasoro algebra representation theory. We will not go into more details here, but only give one example of a generalization of the diamond staggered modules encountered before in~\eqref{stagg-pic-gen-dense-even}, obtained as the scaling limit of a blob algebra modules for $\fug=1$ and $y=1$ (this corresponds to $c=0$ in the CFT language):
 \begin{equation*}
{\scriptsize
   \xymatrix@R=18pt@C=8pt@W=2pt@M=2pt
   { &&&\\
    &&\BW_{j}^{u}\ar[dl]&\\
     &\BW_{j-1}^{b}\ar[dl]& &\\
     \BW_{j-3}^{u} &&&
     } \quad 
      \xymatrix@R=22pt@C=1pt@W=4pt@M=4pt
{&&\\
&&\\
\xrightarrow{\mbox{}\quad N\to\infty\quad}&&\\
&&\\
&&}\quad
   \xymatrix@R=18pt@C=1pt@W=2pt@M=2pt
   { 
    &&h_{1,1-2j}\ar[dl]\ar[dr]&&&\\
     &h_{1,2j-1}\ar[dl]\ar[dr]& &h_{1,5+2j}\ar[dl]\ar[dr]&&\\
     h_{1,7-2j}\ar[dr] &&h_{1,1-2j}\ar[dl]\ar[dr]&&h_{1,-5-2j}\ar[dl]\ar[dr]&\\
          &h_{1,2j-1}\ar[dr]& &h_{1,5+2j} \ar[dl]\ar[dr]&&\quad\ddots\quad\\
               &&h_{1,1-2j}\ar[dr]& & \quad\ddots\quad&\\
                              &&&\quad\ddots\quad &&\\
     }
     }
 \end{equation*}
 Note that this module is a gluing of three Verma modules and it should in particular admit Jordan cells of ranks up to~$3$ for the $L_0$ generator. Finally, the blob algebra admits even more complicated modules -- the tilting modules -- which admit now Jordan cells of any finite rank in the scaling limit~\cite{GJSV}.

\subsection{A remark on a lattice $\mathcal{W}$-algebra}
We can actually go further and introduce even bigger algebras that in the scaling limit give representations of the so-called triplet $\mathcal{W}$-algebra containing Virasoro as a proper subalgebra. These $\mathcal{W}$-algebras are generated by an $SU(2)$ triplet of primary fields,  in addition to the stress-tensor field $T(z)$~\cite{[K-first]}.

 An example of such a construction was given recently~\cite{GST} where lattice regularizations
of the $\mathcal{W}$-algebra generators were proposed. The definition of these lattice $\mathcal{W}$-algebras $\Wlatq{\q}{N}$ is rather technical and they can be roughly described as an extension of the TL algebra by the group $SU(2)$. At the free fermion point, $\q = i$, these algebras are closely connected with known wall Brauer algebras. The scaling limit of lattice models in this case corresponds to symplectic fermions with central charge $c = -2$ and it is straightforward to provide a full analysis of the scaling limit of the lattice Virasoro
and $\mathcal{W}$ algebras, and to show in details how the corresponding continuum Virasoro and $\mathcal{W}$-algebra generators are obtained~\cite{GST}. For higher roots of unity, $\q=\rme^{i\pi/p}$ corresponding to $(p-1,p)$ theories in the scaling limit, the lattice algebras $\Wlatq{\q}{N}$ have tilting modules described in Fig.~\ref{fig:W-proj}, and the number of such modules is not growing with $N$. The structure of these modules does not depend on the number of sites and should thus persist in the scaling limit. The irreducible modules $\modwX^{\pm}_{s}$ in the limit involve an infinite number of Virasoro primary fields while $\modwY_s$ modules correspond to the usual minimal models content. We  note finally that indecomposable  modules over the chiral triplet $\mathcal{W}$-algebra for $c=0$ ($p=3$) with the same subquotient structure were proposed in~\cite{GabRunW2} using  a very different approach (for any $(p-1,p)$ theory similar modules with $5$ and $4$ subquotients involving the minimal model content were constructed in~\cite{[FGST3]}).

\begin{figure}
\begin{equation*}
   \xymatrix@R=14pt@C=12pt@W=3pt@M=2pt{
    &&\stackrel{\modwX^+_{s}}{\bullet}
    \ar@/^/[dl]
    \ar[d]^{}
    \ar@/_/[dr]
    &\\
    &\stackrel{\modwX^-_{p - s}}{\circ}\ar@/^/[dr]
    &\stackrel{\modwY_{s}}{\cdot}\ar[d]^{}
    &\stackrel{\modwX^-_{p - s}}{\circ}\ar@/_/[dl]
    \\
    &&\stackrel{\modwX^+_{s}}{\bullet}&
  }
\quad
\xymatrix@R=14pt@C=12pt@W=3pt@M=2pt{
    &&\stackrel{\modwX^-_{p-s}}{\circ}
    \ar@/^/[dl]
    \ar@/_/[dr]
    &\\
    &\stackrel{\modwX^+_{s}}{\bullet}\ar@/^/[dr]
    &
    &\stackrel{\modwX^+_{s}}{\bullet}\ar@/_/[dl]
    \\
    &&\stackrel{\modwX^-_{p-s}}{\circ}&
  }
\end{equation*}
\caption{Tilting modules of the lattice W-algebra $\Wlatq{\q}{N}$, for $\q=\rme^{i\pi/p}$, $1\leq s\leq p-1$, and $p\geq3$. Its irreducible modules are denoted by $\modwX^{\pm}_s$ and  $\modwY_s$. The same diagrams also describe the subquotient structure of indecomposable modules over the chiral triplet $\mathcal{W}$-algebra from~\cite{[FGST3]}.}
\label{fig:W-proj}
\end{figure}

\section{Indecomposable fusion rules}\label{sec:fusion}

\label{secFusion}

To conclude this review on lattice regularizations of LCFTs, we now describe a procedure allowing to compute fusion rules on the lattice~\cite{RS2,RS3,GV}. The procedure was outlined in \cite{RS2}, and also developed independently --  with, we believe, less algebraic background by Pearce, Rasmussen and Zuber in~\cite{PRZ} and in {\it e.g.}~\cite{RP1,RP2}.

\subsection{Fusion on the lattice and in the continuum}

\begin{figure}
\begin{center}
\psfig{file=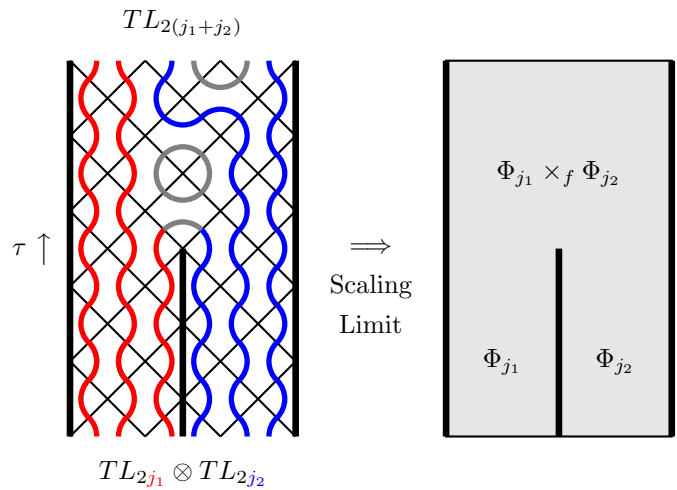}
\end{center}
\caption{Physical interpretation of the lattice fusion of two standard TL modules $\StTL{j_1}[N_1]$ and $\StTL{j_2}[N_2]$ (in the picture, 
  $N_1=2j_1$ and $N_2=2j_2$ so that both standard modules are one-dimensional). Fusion can then be seen as an event in imaginary
  time $\tau$, consisting in ``joining'' the two standard modules by acting with an additional TL generator (induction procedure). In the scaling limit, we expect
  this construction to coincide with the usual fusion procedure or OPE of boundary fields, here $\Phi_{j_1}=\Phi_{1,1+2 j_1}$ and $\Phi_{j_2}=\Phi_{1,1+2 j_2}$, living in the corresponding Virasoro modules.}
\label{aba:figFusion}
\end{figure}

The lattice fusion that we are going to present here was introduced in~\cite{RS2,RS3}, and studied in details in~\cite{GV}.  
The idea is that fusion corresponds to joining two spin chains, each one carrying a representation of the TL algebra, by acting with an additional TL generator at their junction. 
In the scaling limit, those lattice representations will eventually become 
representations of the group of conformal transformations in the interior of the strips. In a more mathematical language, fusion can be thought of as an induction process.
Because of the additional TL generator that will join the two spin chains, or any pair of TL modules, one expects a single copy
of the conformal group to emerge, which contains the tensor product of the conformal groups associated with the two initial strips.
Therefore, the induction process over the Temperley-Lieb algebra corresponds in 
the continuum limit to the induction over the group of conformal transformations in the corresponding regions. Fusion then corresponds to a slit-strip geometry (see Fig.~\ref{aba:figFusion}) that can be 
mapped by a Schwarz--Christoffel transformation~\cite{GSchW} onto the upper half plane, where both sides and the slit of the strip are mapped onto the real line.
Then, the incoming and outcoming states correspond to fields localized at points on the boundary of the half plane.
One can then recover the usual interpretation of the fusion as OPE of the boundary fields.

Formally, the fusion associates with any pair of  modules over the algebras $\TLq{N_1}$ and $\TLq{N_2}$ a module over the bigger algebra $\TLq{N_1+N_2}$. 
Let $M_1$ and $M_2$ be two modules over $\TLq{N_1}$ and $\TLq{N_2}$ respectively, with the same fugacity $\fug$. Then, the tensor product $M_1\tensor M_2$ is a module over the product $\TLq{N_1}\tensor\TLq{N_2}$ of the two algebras. We note that this product of algebras is naturally a subalgebra in $\TLq{N_1+N_2}$.
\textit{The fusion}~$\fus$ of two modules $M_1$ and $M_2$  is then defined as the  module induced from this subalgebra, {\it i.e.} 
\begin{equation}\label{fusfunc-def}
M_1\fus M_2 = \TLq{N_1+N_2}\tensor_{\TLq{N_1}\tensor\TLq{N_2}} M_1\tensor M_2,
\end{equation}
where the balanced product $\tensor_A$ (of right and left modules)  over an algebra $A$ is defined as a quotient of the usual tensor product by the relations $v_1\rightact a\tensor v_2 = v_1\tensor a\leftact v_2$ for all $a\in A$, where the left and right actions of $A$ are denoted by $\leftact$ and $\rightact$, respectively. In simple words, we simply allow any element from $A$  to pass through the tensor-product symbol from right to left and {\it vice versa}. In our context, the algebra $A$ is $\TLq{N_1}\tensor\TLq{N_2}$ and we consider $\TLq{N_1+N_2}$ as a bimodule over itself, with the left and right actions given by the multiplication, and in particular it is a right module over the subalgebra $A$.  The space $M_1\fus M_2$ in~\eqref{fusfunc-def} is then a left module over $\TLq{N_1+N_2}$. 
 For any pair of left modules $M_1$ and $M_2$ over $\TLq{N_1}$ and $\TLq{N_2}$ 
 we shall call \textit{fusion rules} the decomposition of the induced module into indecomposable direct summands.
 
When $\q$ is not a root of unity, it is quite easy to convince oneself that the fusion rules for the TL standard modules 
follow a simple $s\ell(2)$ spin addition rule
\begin{equation}\label{eqFusionGen}
\StTLn{j_1}{N_1}\fus\StTLn{j_2}{N_2} = \bigoplus_{j=|j_1-j_2|}^{j_1+j_2} \StTLn{j}{N_1+N_2},
\end{equation}
for $2j_1 \leq N_1$ and $2j_2 \leq N_2$. This relation to $s\ell(2)$ is not a coincidence of course
and is related by the centralizing property with $\LQG$ with a dual construction~\cite{RS2,RS3,GV} -- the quantum-group fusion. 
 A direct argument for~\eqref{eqFusionGen} is given by considering the geometric interpretation of the induced module $\StTLn{j_1}{N_1=2j_1}\fus\StTLn{j_2}{N_2=2j_2}$ in terms of link diagrams. This module is filtered by  (or composed of)  subspaces indexed by the number $j$ of through-lines which obviously takes integer values from $|j_1-j_2|$ up to $j_1+j_2$. 
 Then, using a semi-simplicity argument we deduce the direct sum decomposition~\eqref{eqFusionGen}. For other values of $N_1$ and $N_2$, the decomposition can be shown in a similar way. We note that a rigorous derivation of the generic fusion~\eqref{eqFusionGen} can be found in section~4 of~\cite{GV}. The TL induction was also recently studied in~\cite{RidoutSaintAubin}.
  
In a language more familiar to physicists, this is reminiscent of the well-known fusion of Kac operators. Indeed, using the correspondence between standard modules and Virasoro Kac modules, this generic fusion corresponds to 
\begin{equation}
\Phi_{1,1+2 j_1}\fus \Phi_{1,1+2 j_2} = \sum_{j=|j_1-j_2|}^{j_1+j_2} \Phi_{1,1+2 j},  
\end{equation}
where $\Phi_{1,1+2 j}$ has conformal weight $h_{1,1+2 j}$.

\subsection{$c \rightarrow 0$ catastrophe on the lattice and OPEs in the continuum limit}

When $\q$ is a root of unity things become much more complicated and one encounters once again indecomposability.
As an example, let us discuss how the $c \rightarrow 0$ catastrophe (see section~\ref{subsecCatastrophe}) manifests itself on the lattice.

Let us consider the fusion $\StTLn{1}{2}\fus\StTLn{1}{2}$,
 where $\StTLn{1}{2}$ has the basis $\{ \ \psset{xunit=2mm,yunit=2mm}
\begin{pspicture}(0,0)(1.5,1)
\psline[linecolor=black,linewidth=1.0pt](0,-0.5)(0,1)
\psline[linecolor=black,linewidth=1.0pt](1,-0.5)(1,1)
\end{pspicture}
\}$ with $e_1 \ \psset{xunit=2mm,yunit=2mm}
\begin{pspicture}(0,0)(1.5,1)
\psline[linecolor=black,linewidth=1.0pt](0,-0.5)(0,1)
\psline[linecolor=black,linewidth=1.0pt](1,-0.5)(1,1)
\end{pspicture}=0$.
The induction results in a six-dimensional $\TLq{4}$-module with the basis
\begin{equation}\label{fus:StTL-3}
\StTLn{1}{2}\fus\StTLn{1}{2} = \langle\, l,\, e_2 l,\, e_1e_2 l,\,  e_3e_2 l,\, e_1e_3e_2 l,\, e_2e_1e_3e_2 l\,\rangle,
\end{equation}
with $l=\psset{xunit=2mm,yunit=2mm}
\begin{pspicture}(0,0)(1.5,1)
\psline[linecolor=black,linewidth=1.0pt](0,-0.5)(0,1)
\psline[linecolor=black,linewidth=1.0pt](1,-0.5)(1,1)
\end{pspicture} \otimes
\psset{xunit=2mm,yunit=2mm}
\begin{pspicture}(0,0)(1.5,1)
\psline[linecolor=black,linewidth=1.0pt](0,-0.5)(0,1)
\psline[linecolor=black,linewidth=1.0pt](1,-0.5)(1,1)
\end{pspicture}$.
This module is decomposed for $\q$ generic as
\begin{equation}\label{fus:StTL-3-gen}
\StTLn{1}{2}\fus\StTLn{1}{2} = \StTL{0}[4] \oplus \StTL{1}[4] \oplus \StTL{2}[4],
\end{equation}
where the two-dimensional invariant subspace $\StTL{0}[4]$ is spanned by $e_1e_3e_2 l$ and $e_2e_1e_3e_2 l$ which may be identified with the link states $\psset{xunit=2mm,yunit=2mm}
\begin{pspicture}(0,0)(3,1)
 \psellipticarc[linecolor=black,linewidth=1.0pt]{-}(0.5,1.0)(0.5,1.42){180}{360}
 \psellipticarc[linecolor=black,linewidth=1.0pt]{-}(2.5,1.0)(0.5,1.42){180}{360}
\end{pspicture}$ and $\psset{xunit=2mm,yunit=2mm}
\begin{pspicture}(0,0)(3,1)
 \psellipticarc[linecolor=black,linewidth=1.0pt]{-}(1.5,1.0)(1.5,1.42){180}{360}
 \psellipticarc[linecolor=black,linewidth=1.0pt]{-}(1.5,1.0)(0.5,0.71){180}{360}
\end{pspicture}$, respectively.
The invariant one-dimensional subspace $\StTL{2}[4]$ is spanned, after solving a simple system of linear equations, by
\begin{equation}\label{fus:StTL-3-inv}
\inv(\fug) = l + \ffrac{1}{\fug^2-2}\Bigl( e_1 e_2l + e_3 e_2 l - \fug e_2 l + \ffrac{1}{\fug^2-1}(e_2e_1e_3e_2 l - \fug e_1e_3e_2 l )\Bigr),
\end{equation}
with $e_j\inv(\fug)=0$, for $j=1,2,3$.
Moreover, three remaining linearly independent states contribute to the three-dimensional irreducible direct summand isomorphic to  $\StTL{1}[4]$ because the algebra is semisimple for generic $\q$. Once again, in terms of Virasoro fields, this generic fusion corresponds to 
\begin{equation}
\Phi_{1,3}\fus \Phi_{1,3} = \Phi_{1,1} + \Phi_{1,3} +\Phi_{1,5},  
\end{equation}
or, more explicitly, 
\begin{multline}\label{eqPercoOPEgeneric}
\displaystyle \Phi_{1,3}(z) \Phi_{1,3}(0) \sim  \frac{1}{z^{2 h_{1,3}}} \left[\one + \frac{2 h_{1,3}}{c} z^2 T(0) + \dots \right] + \frac{C^{\Phi_{1,3}}_{\Phi_{1,3},\Phi_{1,3}}}{z^{h_{1,3}}} \left[\Phi_{1,3}(0)+ \frac{z}{2} \partial \Phi_{1,3}(0) +\dots \right] \\ + \frac{C^{\Phi_{1,5}}_{\Phi_{1,3},\Phi_{1,3}}}{z^{2 h_{1,3}-h_{1,5}}} \left[\Phi_{1,5}(0) + \frac{z}{2} \partial \Phi_{1,5}(0) +\dots \right].
\end{multline}

We see that the submodules $\StTL{0}[4]$ and $\StTL{1}[4]$ (or their basis elements) have a well-defined limit $\fug\to1$ ($p=3$, percolation) while the invariant $\inv(\fug)$ spanning $\StTL{2}[4]$ is not defined in this limit -- the state in~\eqref{fus:StTL-3-inv} has a term diverging as $\fug\to1$. As it turns out, this can be thought of as the lattice analog of the $c \rightarrow 0$ catastrophe in the OPE~\eqref{eqPercoOPEgeneric}. The resolution of this lattice catastrophe was discussed in details in~\cite{GV}. The idea is to introduce the new state
\begin{equation}
t(\fug) = \inv(\fug) - \ffrac{1}{(\fug^2-2)(\fug^2-1)}\bigl(e_2e_1e_3e_2 l + a_- e_1e_3e_2 l\bigr),
\end{equation}
with  $a_{-}=-h_{-}(\fug)-\fug$ and $h_{-}(\fug) = -\frac{3\fug - \sqrt{8+\fug^2}}{2}$. It can be easily shown that the state $t(\fug)$ has a finite limit
as $\fug \rightarrow 1$. Borrowing the terminology of LCFT, we say that the state $t$ 
is the ``logarithmic partner'' of the ``stress-energy tensor'' $T=\psset{xunit=2mm,yunit=2mm}\begin{pspicture}(0,0)(3,1)
 \psellipticarc[linecolor=black,linewidth=1.0pt]{-}(0.5,1.0)(0.5,1.42){180}{360}
 \psellipticarc[linecolor=black,linewidth=1.0pt]{-}(2.5,1.0)(0.5,1.42){180}{360}
\end{pspicture}$ \ - \ $\psset{xunit=2mm,yunit=2mm}
\begin{pspicture}(0,0)(3,1)
 \psellipticarc[linecolor=black,linewidth=1.0pt]{-}(1.5,1.0)(1.5,1.42){180}{360}
 \psellipticarc[linecolor=black,linewidth=1.0pt]{-}(1.5,1.0)(0.5,0.71){180}{360}
\end{pspicture}$. Indeed, we find a Jordan cell between these two states
\begin{equation}
H t = \ffrac{2}{3}T.
\end{equation}
We will also say that $T$ is the ``descendant'' of the vacuum state $\vacr = \psset{xunit=2mm,yunit=2mm}
\begin{pspicture}(0,0)(3,1)
 \psellipticarc[linecolor=black,linewidth=1.0pt]{-}(1.5,1.0)(1.5,1.42){180}{360}
 \psellipticarc[linecolor=black,linewidth=1.0pt]{-}(1.5,1.0)(0.5,0.71){180}{360}
\end{pspicture} \ +2 \ \psset{xunit=2mm,yunit=2mm}
\begin{pspicture}(0,0)(3,1)
 \psellipticarc[linecolor=black,linewidth=1.0pt]{-}(0.5,1.0)(0.5,1.42){180}{360}
 \psellipticarc[linecolor=black,linewidth=1.0pt]{-}(2.5,1.0)(0.5,1.42){180}{360}
\end{pspicture}$ as the standard module $\StTL{0}$ has the following indecomposable structure at $\fug=1$:
$\StTL{0} = \vacr \to T$ where we recall that the arrow corresponds to the action of the TL algebra.

We see that the standard modules $\StTL{0}[4]$ and $\StTL{2}[4]$ arising
in the generic fusion rules are ``glued'' together at $\fug=1$
into a bigger indecomposable module with the TL action given by the diagram $t\to\vacr\to T$. The subquotient structure of this module reads  $\IrrTL{2}\to\IrrTL{0}\to\IrrTL{2}$, where each subquotient is one-dimensional and we recall that  $\IrrTL{j}$ denotes the irreducible  top of $\StTLn{j}{N}$. We will denote the resulting module $\PrTL{2}[4]$; this is an example of tilting module (see section~\ref{subsecRepTheoryTL}). Finally, the fusion rules at $\fug=1$ read
\begin{equation}\label{fus:StTL-3-perc}
\StTLn{1}{2}\fus\StTLn{1}{2} =  \StTL{1}[4] \oplus \PrTL{2}[4],\qquad \text{for}\; p=3.
\end{equation}
In the scaling limit, this means that the stress energy tensor $T = L_{-2} \one$ is mixed with $\Phi_{1,5}$ at $c=0$ into the new field $t$.
Just like we did on the lattice, one can introduce a new field $t(z)$ for generic central charge as
\begin{equation}\label{eqDeftOPE}
\displaystyle  t (z)  = C^{\Phi_{1,5}}_{\Phi_{1,3},\Phi_{1,3}} \ffrac{b(\epsilon)}{h_{1,3}} \Phi_{1,5} (z)+ \ffrac{b(\epsilon)}{\Braket{T | T}} T(z),
\end{equation}
where $b(\epsilon) = - \frac{ \Braket{T | T}}{h_{1,5}-2}$, $\Braket{T | T}=\frac{c}{2}$ and $p=3+\epsilon$. 
The OPE~\eqref{eqPercoOPEgeneric} has then a finite limit
\begin{equation}
\displaystyle \Phi_{1,3}(z) \Phi_{1,3}(0) \sim \frac{1}{z^{2/3}} \left[\one + \ffrac{ 1}{3 b} z^2( T(0) \ln z + t(0)) + \dots \right] + \dots,
\end{equation}
with $b= \lim_{\epsilon \rightarrow 0} b(\epsilon)=-\frac{5}{8}$. Of course, one can also compute correlation functions of $t (z)$ to check that it indeed corresponds to the logarithmic partner of $T(z)$. 

This is just one example of a lattice fusion rule, in good agreement with what is expected on the field theory side. Using the bimodule structure of the spin chains  and algebra involving quantum group results, it is actually possible to 
obtain rigorous, general results for the lattice fusion of most of Temperley-Lieb modules~\cite{GV} for all roots of unity $\q$. The physical consequences for the OPEs were also discussed in~\cite{GV}.

\section{Periodic models and bulk LCFTs}
While the case of boundary LCFTs is thus slowly getting under control,
the understanding of the {\sl bulk} case remains in its infancy. The
main problem here, from the continuum point of view, is the expected
double indecomposability of the modules
over the product of the left and right Virasoro algebras, leading to
potentially very complicated modules which have proven too hard to
study so far, except in some special cases. These include bulk
logarithmic theories~\cite{GabRun,GabRunW} with $\mathcal{W}$-algebra
symmetries~\cite{[K-first],[FGST3]}, and WZW models on
supergroups which, albeit very simple as far as LCFTs go, provide
interesting lessons on the coupling of left and right
sectors~\cite{SchomSal, SaleurSchomerus}. 

 From the lattice point of view, the necessarily periodic geometry of the model
leads to more complicated algebras~\cite{MartinSaleur1, MartinSaleur},
like affine or periodically extended TL algebras,
and to a more intricate role of the quantum
group~\cite{PasquierSaleur}, whose symmetry is partly lost. While it is
possible to define and study lattice models whose continuum limit is a
(bulk) LCFT, the underlying structures are also very difficult to get:
the lattice algebras have a much more
complicated representation theory. Nevertheless, it looks possible
to generalize the approach discussed above in the context of boundary spin-chains, 
partly to make progress in abstract representation theory of the periodic lattice algebras~\cite{GL}
and to obtain some results on their full symmetry algebras~\cite{GRS1}.

At least in free fermion cases, the lattice analysis given in~\cite{GRS1}
shows how one can straightforwardly  proceed from the periodic Temperley--Lieb (PTL)
generators to get  Virasoro modes in the bulk LCFT of symplectic fermions:
 the combinations 
\begin{equation}\label{HPn-def}
H(n) = -\sum_{j=1}^N e^{-iqj} e_j,\qquad
P(n)=\ffrac{i}{2}\sum_{j=1}^N e^{-iqj} [e_j,e_{j+1}], \qquad q=\ffrac{2n\pi}{N},
\end{equation}
of the
PTL generators $e_j$  converge in the scaling limit as $N\to\infty$ to the well-known symplectic fermions representation
of the left and right Virasoro generators
\begin{equation*}
\ffrac{N}{4\pi}H(n) \mapsto 
L_{n}+\bar{L}_{-n}, \qquad \ffrac{N}{4\pi}P(n) \mapsto 
L_{n}-\bar{L}_{-n},
\end{equation*}
where the limit is
taken for finite $n$. 

In general, while we expect to be able to extract the stress-energy
tensor modes, $L_n$ and $\bar{L}_n$, from the periodic
Temperley--Lieb algebras~\cite{KooSaleur,GRS1}, the interesting point is that the scaling limit of 
 elements in the periodic TL can lead to other physical observables corresponding
to different bulk scaling fields.  A very important example of
such a field is the energy operator\footnote{This is the field canonically coupled to the temperature.} in the Potts model, associated
with the staggered sum
\begin{equation}\label{energypotts}
\sum_{i=1}^{N} (-1)^i e_i\mapsto \int {\rm d}x\, \Phi_{2,1}\times \overline{\Phi}_{21}(x,\tau=0)
\end{equation}
where the integral is taken over the
  circumference of a cylinder at constant imaginary-time 
  $\tau=0$. The field  in~\eqref{energypotts} is the non-chiral degenerate field with conformal weights $h=\bar{h}=h_{2,1}$.  Of course, the introduction of such fields in the organizing
  algebra of a LCFT requires discussion of objects which mix chiral
  and anti-chiral sectors. This leads us to the new
  concept of {\it interchiral algebra}~\cite{GRS3}. 
  
For periodic $g\ell(1|1)$ spin-chains or free fermion points of XXZ models
with some twists, an exhaustive analysis was done in~\cite{GRS3} where the structure
  and role of the interchiral algebra in the case of  the bulk symplectic fermions was also discussed in details.
The idea there was to   
   consider also limits
with $n$ close to $N/2$ (the dispersion relation of our Hamiltonian has low-energy part close to zero  as well as to $\pi/2$ momenta).
It turns out that this limit
gives the modes of 
a field  $\enrg(z,\bar{z})$  (of conformal dimension $(1,1)$) which is
expressed in terms of  derivatives of the symplectic fermions $\Phi^{\alpha}$ as
\begin{equation}\label{interfielddef}
\enrg(z,\bz)=S_{\alpha\beta}\psi^{\alpha}(z)\bar{\psi}^{\beta}(\bz)\qquad \text{with}\quad
\psi^{\alpha}(z)=\der\Phi^{\alpha}(z,\bz),\quad
\bar{\psi}^{\alpha}(\bar{z})=\bder\Phi^{\alpha}(z,\bz),
\end{equation}
where we use the symmetric form 
\begin{equation}
 S_{12}=S_{21}=1, \quad
S_{11}=S_{22}=0.
\end{equation}
Remarkably, this additional field  generates the
 full scaling limit of the PTL algebras (represented in the $\gl(1|1)$ spin chains) and this limit is our interchiral algebra $\interchalg$.
This statement was actually proven~\cite{GRS3} using an interesting connection with symplectic Lie algebras
$\ssp_{N-2}$ (a Howe duality) and their rigorous direct limit $\spinf$.
In particular, on the finite lattice with $N$ sites the PTL simple modules are just fundamental representations of $\ssp_{N-2}$, in the limit simple modules over $\interchalg$ are identified with appropriate simple $\spinf$-modules.

\section{Beyond two dimensions}

To this point we have extensively exposed ideas and tools that are proper
to two dimensions. However, the fundamental mechanism for producing Jordan cells of
the dilatation operator remains operative in higher dimensions, $d > 2$, provided
that two (or more) suitably related operators possess coinciding scaling dimensions.
The algebraic tools that would permit to compute the ensuing logarithmic structure
directly within such an LCFT are however missing. Instead, insight can be gained by
accessing that theory as a limit, by tuning a suitable continuous (or formally
continuous) parameter. This point of view is discussed in the contribution of Cardy
\cite{CardyThis} to this Special Issue.

Several situations of this kind are of
direct physical relevance:
\begin{enumerate}
 \item Disordered systems described by $N$-fold replication,
 in the replica limit $N \to 0$ \cite{CardyLCFT,Cardy01};
 \item The O($n$) model, in the limit $n \to$ a non-positive integer 
 (including the polymer limit $n \to 0$) \cite{CardyLCFT}; 
 \item The $Q$-state Potts model, in the limit $Q \to$ a
 non-negative integer (including the bond-percolation limit $Q \to 1$)
 \cite{CardyLCFT,VJS12,VJS13}.
\end{enumerate}
In all cases the key assumption is that physical
operators can be fully described as irreducibles of the corresponding symmetry group
($S_N$, O($n$) or $S_Q$, as the case may be). Obviously, this approach will fail to
give exhaustive results if the actual symmetry turns out to be larger (e.g., when
specializing the results for the Potts model in general dimension to $d=2$).
This line of research has been pushed the furthest for the Potts model
\cite{VJS12,VJS13} and we briefly review some of the results obtained.

The starting point is to define $S_Q$ symmetric tensors of rank $k \le N$ acting on $N$ lattice spins,
subject to the constraints that 1) the tensor vanishes unless all $N$ spins are distinct
and 2) the sum over any of the $k$ tensor indices is zero. For $k=N=1$ this obviously
produces $\varphi_a = \delta_{\sigma_i,a} - 1/Q$ which is just the Potts order parameter
(magnetization operator). For $N=2$ one finds \cite{VJS12}
\begin{subequations}
\label{eqPott2Spin}
\begin{eqnarray}
 E(\sigma_i,\sigma_j) &=& \delta_{\sigma_i \neq \sigma_j}  \,, \\
 \phi_a(\sigma_i,\sigma_j) &=&  \delta_{\sigma_i \neq \sigma_j} \left( \varphi_a(\sigma_i)+\varphi_a(\sigma_j) \right)   \,, \\
 \hat{\psi}_{ab} (\sigma_i,\sigma_j)&=& \delta_{a \neq b} \left( \delta_{\sigma_i, a} \delta_{\sigma_j, b} +\delta_{\sigma_i, b} \delta_{\sigma_j, a} -\frac{\phi_a+ \phi_b}{Q-2}   -\frac{2E}{Q(Q-1)} \right) \,. \label{hatpsi}
\end{eqnarray} 
\end{subequations}
The poles at $Q=0,1,2$ are indicative of special behavior at this integers. More generally,
one can construct such tensors for arbitrary $k$ and $N$, in which case poles are found at all non-negative
integers $Q$ \cite{VJS13}.

We focus henceforth on percolation ($Q \to 1$) in the bulk. A well-defined limit is obtained by mixing
$\hat{\psi}_{ab}$ with $E$. The energy operator is given by 
$\varepsilon (r_i) \equiv E(\sigma_i,\sigma_j) - \langle E \rangle$, where we have taken the scaling
limit with $j=i+1$. For generic (real) $Q$ one has
\begin{equation}
\displaystyle \langle \varepsilon (r)  \varepsilon (0) \rangle = \tilde{A}(Q) (Q-1) r^{-2 \Delta_{\varepsilon}(Q)} \,,
\end{equation} 
where $\tilde{A}(Q)$ is a regular function of $Q$, with a finite non-zero limit $\tilde{A}(1)$ for $Q \to 1$.
The factor $(Q-1)$ follows from the general result \cite{Flohrreview,VGJS} that correlation functions containing
only energy operators vanish in bulk percolation.
The tensor structure of two-point functions of arbitrary tensors can be deduced by combinatorial means \cite{VJS13}.
In particular one has \cite{VJS12}
\begin{multline}
\label{eqTwoPointpsiHat}
\langle \hat{\psi}_{ab} (r) \hat{\psi}_{cd} (0) \rangle = \frac{2 A(Q)}{Q^2} \left( \delta_{ac} \delta_{bd} + \delta_{ad} \delta_{bc} - \frac{1}{Q-2} \left(\delta_{ac}+\delta_{ad}+\delta_{bc}+\delta_{bd} \right)  \right. \\
+\left. \frac{2}{(Q-1)(Q-2)} \right)
 \times r^{-2 \Delta_{\hat{\psi}}(Q)},
\end{multline} 
where $A(Q)$ is again a regular function of $Q$ when $Q\rightarrow 1$. Moreover, for the maximum
rank $k=N$ the two-point function is proportional to the probability of having $N$ distinct clusters
propagating between the insertion points $0$ and $r$ \cite{VJS13}. In particular, $\Delta_{\hat{\psi}}(Q)$
is the scaling dimension of the 2-cluster operator.

Since $\Delta_{\hat{\psi}}(Q)$ and $\Delta_{\varepsilon}(Q)$ collide for $Q \to 1$ in any dimension $d \ge 2$
\cite{Coniglio82}, it is possible for the operators $\hat{\psi}$ and $\varepsilon$ to mix and form a Jordan cell. Moreover,
they {\em must} do so in order to cure the divergence manifest in the last term of (\ref{eqTwoPointpsiHat}).
The appropriate mixed operator is
$\tilde{\psi}_{ab}  (r) = \hat{\psi}_{ab}  (r) + \frac{2 \varepsilon (r)}{Q(Q-1)}$, with $a \neq b$, and its two-point
function is non-singular in the $Q \to 1$ limit provided that $A(1)=\tilde{A}(1)$:
\begin{equation}
\label{eq_PercoLog}
 \langle \tilde{\psi}_{ab}  (r) \tilde{\psi}_{cd}  (0) \rangle = 2 A(1) r^{- 2 \Delta_{\hat{\psi}}(1)} \left[
 \left( \delta_{ac}+\delta_{ad}+\delta_{bc}+\delta_{bd} + \delta_{ac} \delta_{bd} + \delta_{ad} \delta_{bc} \right) +
 4 K \log r \right] \,.
\end{equation} 
The constant $K = \lim_{Q \rightarrow 1} (\Delta_{\hat{\psi}} - \Delta_{\varepsilon})/(Q-1)$ is universal, but obviously
cannot be computed exactly for $d>2$ due to our ignorance about the scaling dimensions. For $d=2$ one finds
$K = \sqrt{3}/\pi$.

The logarithmic behavior can be isolated by constructing an appropriate geometrical observable \cite{VJS12}.
For $k=0,1,2$, let ${\mathbb P}_k$ be the probability that $k$ distinct clusters join the group of two spins at point $0$ to
the group of two spins at point $r$; and let ${\mathbb P}_{\neq}$ denote the probability that the two spins within a group are distinct.
Then the combination
\begin{equation}
\label{eqLogGeomPerco}
 \displaystyle F(r) \equiv \dfrac{{\mathbb P}_0(r) + {\mathbb P}_1(r) - {\mathbb P}_{\neq}^2}
 {{\mathbb P}_2(r)} \sim 2 K \log r + {\rm const}
\end{equation}
isolates a {\rm pure} logarithmic scaling (i.e., not multiplied by any power law) which is observable in numerical simulations.
For instance, one finds $2K = 1.15 \pm 0.05$ in $d=2$ \cite{VJS12}, which compares
favorably to the exact result $2K = 2\sqrt{3} / \pi \simeq 1.1026$.
Note also that the logarithm has its root in the {\em disconnected} correlation function ${\mathbb P}_0$;
this is true more generally \cite{CardyLCFT}.

Going beyond this specific example, the whole logarithmic structure of two- and
three-point correlation functions in $d > 2$ can be systematically unearthed by constructing
the $S_Q$ irreducible tensors acting on $N$ Potts spins \cite{VJS13}. In some sense, this extends to the realm of LCFT
Polyakov's well-known classification of two- and three-point functions in $d$
dimensions, obtained in his case by using only global conformal invariance
\cite{Polyakov70}. 

\section{Conclusion}

By this direct lattice approach, LCFTs of physical interest are slowly getting under control. The basic relationship between the Virasoro and the Temperley Lieb algebra remains  however, ill understood -- although it is clear it has to do with centralizers~\cite{PasquierSaleur}. While associative algebras provide a natural way to understand and classify Virasoro modules, it is likely that the analysis can be carried out directly at the level of Virasoro -- an infinite dimensional Lie algebra -- provided more mathematical tools are developed. It is also worth emphasizing that the relationship with physics is growing. Apart from the statistical mechanics problems already mentioned, the sigma models appearing on the AdS side of the AdS/CFT correspondence, for instance, are close cousins of the supergroup sigma models discussed here. There has also been interest recently in studying what occurs `beyond the topological sector' in sigma models and Yang Mills theories~\cite{Nekrasov}, {\it etc}. We should also mention that besides concrete physical applications, LCFTs attracted recently considerable interest in the mathematics community~\cite{[HLZ],[AM1],[TW]} as well. In particular, the existence and uniqueness of logarithmic OPEs (including the associativity condition) were actually stated, as a theorem, in the fundamental series of papers~\cite{[HLZ]} (see also the review of Huang and Lepowsky in this volume), many rigorous results on $\mathcal{W}$-algebras are given in~\cite{[AM1],[TW]}, {\it etc}.

Among the remaining challenges, the understanding of bulk LCFTs is of utmost interest. In the lattice approach, bulk LCFTs are tackled by considering periodic spin chains. Because of the `loops' then going around the space direction, the corresponding algebras are considerably more complicated -- see~\cite{GRS1,GRS2,GRS3} for details.  The spectrum of conformal weights is then extremely rich, in particular, the conformal weights cannot be arranged in a finite number of families where weights differ from each other by integers. Although the conformal weights are all rational, the theory is therefore {\sl not rational}~\cite{MooreSeiberg}. The full analysis of the bulk percolation LCFT will be described in details in~\cite{GRSV}, one of the main results being that the Hamiltonian operator $L_0+\bar{L}_0$ admits Jordan cells of arbitrarily large rank as the corresponding conformal weight is increased.

\label{sec:disc}

\section*{Acknowledgments}

We thank our collaborators on these and related topics: R. Bondesan, C. Candu, J. Dubail, Y. Ikhlef, and especially V. Schomerus and  I.Yu. Tipunin.  The work described was supported by the Agence Nationale de la Recherche (grant ANR-10-BLAN-0414: DIME). The work of A.M.G. was supported in part by Marie-Curie IIF fellowship and the RFBR grant 10-01-00408. The work of
N.R. was supported by the NSF grants DMR-0706195 and DMR-1005895.



%

\begin{thebibliography}{99}
\bibitem{BPZ} A.A. Belavin, A.M. Polyakov, A.B. Zamolodchikov, {\it Infinite conformal symmetry in two-dimensional quantum field theory}, Nucl. Phys. B {\bf 241} , 333 (1984).

\bibitem{FQS} D. Friedan, Z. Qiu and S. Shenker, {\it Conformal Invariance, Unitarity, and Critical Exponents in Two Dimensions}, Phys. Rev. Lett. {\bf 52} , 1575 (1984).

\bibitem{Cardy} J. Cardy, {\em Operator content of two-dimensional conformally invariant theories},  Nucl. Phys. {\bf 270}, 186  (1986).

\bibitem{Zamo} A. B. Zamolodchikov, {\em Conformal symmetry and multicritical points in two-dimensional quantum field theory}, Sov. J. Nucl. Phys.{\bf 44} , 529 (1986).

\bibitem{Cardybdr} J. Cardy, {\em Conformal invariance and surface critical behavior }, Nucl. Phys. {\bf B240}, 514 (1984).

\bibitem{AffLud} I. Affleck and A. W. W. Ludwig, {\em The Kondo effect, conformal field theory and fusion rules}, Nucl. Phys. {\bf 352}, 849 (1991).

\bibitem{Delft} J.von Delft, D. C. Ralph, R. A. Buhrman, S. K. Upadhyay, R. N. Louie, A. W. W. Ludwig, V. Ambegaokar,  {\em The 2-Channel Kondo Model I: Review of Experimental Evidence for its Realization in Metal Nanoconstrictions}, Ann. Phys.,{\bf  Vol 263}, 1 (1998).

\bibitem{Saleur87} H. Saleur, {\em Conformal invariance for polymers and percolation}, J. Phys. A {\bf 20}, 455 (1987).

\bibitem{Nienhuis} B. Nienhuis, {\em Critical behavior in two dimensions and charge asymmetry of the Coulomb gas}, J. Stat. Phys. {\bf 34} , 731 (1984).

\bibitem{CardyYL} J. Cardy, {\em Conformal Invariance and the Yang-Lee Edge Singularity in Two Dimensions},  Phys. Rev. Lett. {\bf 54} 1354 (1985).


\bibitem{Klitzing} K.v. Klitzing, G. Dorda, and M. Pepper, {\em New Method for High-Accuracy Determination of the Fine-Structure Constant Based on Quantized Hall Resistance}, Phys. Rev. Lett. {\bf 45}, 494--497 (1980).    

\bibitem{Tsui} D.C. Tsui, H.L. Stoermer and A.C. Gossard, {\em Two-Dimensional Magnetotransport in the Extreme Quantum Limit}, Phys. Rev. Lett. {\bf 48}, 1559  (1982).

\bibitem{Pruisken} H. Levine, S.B. Libby and A.M.M. Pruisken, {\em Theory of the quantized Hall effect}, Nucl. Phys. {\bf B240}, 30 (1984).

\bibitem{Efetov} K. B. Efetov, {\em Supersymmetry and theory of disordered metals}, Adv. Phys. {\bf 32},  53 (1983).


\bibitem{Zirnbauer} M. Zirnbauer, {\it Conformal field theory of the integer quantum Hall plateau transition}, {\tt arXiv:hep-th/9905054} (unpublished), and references therein.

\bibitem{SQHE}  I.A. Gruzberg, A.W.W. Ludwig and N. Read, {\em Exact exponents for the spin quantum Hall transition}, Phys. Rev. Lett. {\bf 82}, 4524 (1999).

\bibitem{RS1} N. Read and H. Saleur, {\em Exact spectra of conformal supersymmetric nonlinear sigma models in two dimensions}, Nucl. Phys. B {\bf 613}, 409 (2001).

\bibitem{Bender} C. Bender, S. Boettcher, P. Meisinger, {\em PT symmetric quantum mechanics}, J.Math.Phys. {\bf 40}, 2201  (1999) .

\bibitem{Forests} S. Caracciolo, J. L. Jacobsen, H. Saleur, A. D. Sokal, and A. Sportiello,  Phys.Rev.Lett. 93 (2004) 080601.

\bibitem{Gurarie} V. Gurarie, {\em Logarithmic operators in conformal field theory}, Nucl. Phys. B {\bf 410} , 535 (1993).


\bibitem{Flohrreview} M. Flohr, {\em Bits and Pieces in Logarithmic Conformal Field Theory}, Int. J. Mod. Phys. {\bf A18}, 4497 (2003).

\bibitem{Gaberdielreview} M. Gaberdiel, {\em An algebraic approach to logarithmic conformal field theory},  	Int. J. Mod. Phys. {\bf A18}, 4593 (2003).

\bibitem{Gotz} G. Gotz, T. Quella and V. Schomerus, {\em Representation theory of $sl(2|1)$},  J.Algebra {\bf 312}, 829 (2007).

\bibitem{RozSal} L. Rozansky and H. Saleur, {\em Quantum field theory for the multi-variable Alexander-Conway polynomial}, Nucl. Phys. B {\bf 376} , 461 (1992).

\bibitem{SchomSal} V. Schomerus and H. Saleur, {\em The $GL(1|1)$ WZW-model: From supergeometry to logarithmic CFT}, Nucl. Phys. B {\bf 734}, 221 (2006).

\bibitem{Eggert} S. Eggert, {\em Numerical evidence for multiplicative logarithmic corrections from marginal operators}, Phys. Rev. B {\bf 54} , 9612 (1996).

\bibitem{CardyThis} J.\ Cardy, {\em Logarithmic conformal field theories as limits of ordinary CFTs and some physical applications}, {\tt arXiv:1302.4279}, in this Special Issue.

\bibitem{CardyLCFT} J.\ Cardy, {\em Logarithmic Correlations in Quenched Random Magnets and Polymers}, {\tt arXiv:cond-mat/9911024}.

\bibitem{Gurariecthm} V. Gurarie, {\em c-Theorem for Disordered Systems}, Nucl.Phys. {\bf B546}, 765  (1999).

\bibitem{FlohrAML} M. Flohr, A. Mueller-Lohmann, {\em Notes on non-trivial and logarithmic CFTs with c=0},  J.Stat.Mech.0604:P04002 (2006)

\bibitem{[FGST3]} B.L.~Feigin, A.M.~Gainutdinov, A.M.~Semikhatov, and
  I.Yu.~Tipunin, \textit{Logarithmic extensions of minimal models:
    characters and modular transformations}, Nucl. Phys. B757 (2006)
  303-343, {\tt hep-th/0606196}.


\bibitem{MathieuRidout} P.~Mathieu and D.~Ridout, {\it From Percolation to Logarithmic Conformal Field Theory}, Phys. Lett. B {\bf 657}, 120 (2007). 

\bibitem{GabRun} M. Gaberdiel and I. Runkel,  \textit{From boundary to
bulk in logarithmic CFT}, J. Phys. A41 (2008) 075402.

\bibitem{GabRunW2} M.\ Gaberdiel, I.\ Runkel and S.\ Wood,
  {\em Fusion rules and boundary conditions in the $c=0$ triplet model},
  J.\ Phys.\ A: Math.\ Theor.\ {\bf 42}, 325403 (2009).


\bibitem{GabRunW} M. R. Gaberdiel, I. Runkel, S. Wood, {\em A modular invariant bulk theory for the c=0 triplet model},   J.Phys. {\bf A44}, 015204 (2011)

\bibitem{PasquierSaleur} V.~Pasquier and H.~Saleur, {\it Common structures between finite systems and conformal field theories through quantum groups}, Nucl. Phys. B {\bf 330}, 523 (1990).

\bibitem{QG-book} C. Gomez, M. Ruiz-Altaba, G. Sierra, \textit{Quantum Groups in Two-Dimensional Physics},
Cambridge University Press, 1996, 475 pp.

\bibitem{M1} P.P.~Martin and D.~McAnally, \textit{On Commutants, Dual Pairs and Non-Semisimple Algebras from Statistical Mechanics}, Int. J. Mod. Phys. A 7, Supp. 1B, 675 (1992); P.P.~Martin, \textit{On Schur--Weyl duality, $A_n$ Hecke algebras and quantum $sl(N)$ on $\otimes^{n+1}{\mathbb C}^N$}, Int. J. Mod. Phys. A {\bf 7}, Supp. 1B, 645 (1992).

\bibitem{[FGST2]} B.L.~Feigin, A.M.~Gainutdinov, A.M.~Semikhatov, and
  I.Yu.~Tipunin, \textit{Kazhdan--Lusztig correspondence for the
    representation category of the triplet $W$-algebra in logarithmic
    CFT}, Theor.\ Math.\ Phys.\ {\bf 148}  1210--1235 (2006).

\bibitem{[FGST4]} B.L. Feigin, A.M. Gainutdinov, A.M. Semikhatov, and I.Yu. Tipunin, \textit{Kazhdan--Lusztig-dual quantum group for logarithmic extensions of Virasoro minimal models}, J. Math. Phys. {\bf 48} 032303 (2007).


\bibitem{[GT]}    A.M. Gainutdinov and I.Yu. Tipunin, \textit{Radford, Drinfeld, and Cardy boundary states in $(1, p)$ logarithmic conformal field models}, J. Phys. A: Math. Theor. {\bf 42} 315207 (2009).

\bibitem{[BGT]} P.V. Bushlanov, A.M. Gainutdinov, and I.Yu. Tipunin, \textit{Kazhdan-Lusztig equivalence 
and fusion of Kac modules in Virasoro logarithmic models},  Nucl. Phys. B {\bf 862} 232--269 (2012).

\bibitem{ST2} A.M. Semikhatov and I.Yu. Tipunin,
\textit{Logarithmic $\hat{sl}(2)$ CFT models from Nichols algebras. 1}, {\tt arXiv:1301.2235}.
    

\bibitem{AubinLoop} A. Morin-Duchesne and Y. Saint-Aubin, {\em The Jordan Structure of Two Dimensional Loop Models},  	J.Stat.Mech.1104:P04007, (2011).

\bibitem{AubinLoopThis} A. Morin-Duchesne and Y. Saint-Aubin, {\em Jordan cells of periodic loop models}, {\tt arXiv:1302.5483}, (2013).

\bibitem{PRZ} P.~Pearce, J.~Rasmussen and J.B.~Zuber, \textit{Logarithmic minimal models}, J. Stat. Mech. P11017 (2006). 

\bibitem{RP1} J.~Rasmussen and P.~Pearce, {\it Fusion algebra of critical percolation}, J. Stat. Mech. P09002 (2007). 

\bibitem{RP2} J.~Rasmussen and P.~Pearce, {\it Fusion Algebras of Logarithmic Minimal Models}, J. Phys.  A {\bf 40}, 13711 (2007). 

\bibitem{PRV} P. A. Pearce, J. Rasmussen, and S. P. Villani,
\textit{Infinitely extended Kac table of solvable critical dense polymers}, {\tt arXiv:1210.8301}.
    

\bibitem{GV} A.M.\ Gainutdinov and R.\ Vasseur, {\it Lattice fusion rules and logarithmic operator product expansions}, Nucl. Phys. B {\bf 868}, 223--270, (2013) .

\bibitem{RidoutSaintAubin} D.\ Ridout and Y.\ Saint-Aubin,
 {\em Standard modules, induction and the Temperley-Lieb algebra}, {\tt arXiv:1204.4505}. 
 

\bibitem{M0} P.P.~Martin, \textit{Potts models and related problems in statistical mechanics}, World Scientific (1991). 


\bibitem{ReadSaleur01} N.~Read and H.~Saleur, {\it Exact spectra of conformal supersymmetric nonlinear sigma models in two dimensions}, Nucl. Phys. B {\bf 613}, 409 (2001). 

\bibitem{RS2}  N.\ Read and H.\ Saleur, {\em Enlarged symmetry algebras of spin chains, loop models, and S-matrices},
 Nucl.\ Phys.\ B {\bf 777}, 263--315 (2007). 

\bibitem{GRS1} A.M.~Gainutdinov, N.~Read and H.~Saleur,
  \textit{Continuum limit and symmetries  of  the periodic $\gl(1|1)$
    spin chain}, {\tt arXiv:1112.3403}.
    

\bibitem{GRS2} A.M.\ Gainutdinov, N.\ Read and H.\ Saleur, {\em Bimodule structure in the periodic $\gl(1|1)$ spin chain},
 {\tt arXiv:1112.3407}.

\bibitem{GRS3} A.M.\ Gainutdinov, N.\ Read and H.\ Saleur, {\em Associative algebraic approach to  logarithmic CFT in the bulk: 
 The continuum limit of the $\gl(1|1)$ spin chain and the interchiral algebra}, {\tt arXiv:1207.6334}. 
     

\bibitem{GRSV} A.M.\ Gainutdinov, N.\ Read, H.\ Saleur and R.\ Vasseur, to appear.

\bibitem{Westbury} B.W.~Westbury, \textit{The representation theory of the Temperley--Lieb algebras}, Math. Z. 219, 539--565 (1995).

\bibitem{MWood} P.P.~Martin and D.~Woodcock, {\it On quantum spin-chain spectra and the structure of Hecke algebras and q-groups at roots of
unity}, J. Phys. A: Math. Gen. {\bf 31}, 10131 (1998).

\bibitem{RidS}  D. Ridout, Y. Saint-Aubin, 
\textit{Standard Modules, Induction and the Temperley-Lieb Algebra}, arXiv:1204.4505.


\bibitem{Donkin} S.~Donkin, \textit{The q-Schur Algebra},  
London Mathematical Society Lecture Note Series, 1998.
  

\bibitem{GJSV} A.M. Gainutdinov, J.L. Jacobsen, H. Saleur, R. Vasseur, \textit{A physical approach to the classification of indecomposable Virasoro representations from the Blob algebra}, Nucl. Phys. B {\bf 873}, 614--681 (2013).

\bibitem{AF-book} F.W. Anderson and K.R. Fuller, \textit{Rings and Categories of Modules}, Graduate Texts in Math. 13,
Springer-Verlag, NY, 1992.

\bibitem{RS3} N.\ Read and H.\ Saleur, {\em Associative-algebraic approach to logarithmic conformal field theories},
 Nucl.\ Phys.\ B {\bf 777}, 316--351 (2007). 

\bibitem{KooSaleur} W.M.~Koo and H.~Saleur, {\it Representations of the Virasoro algebra from lattice models}, Nucl. Phys. B {\bf 426}, 459 (1994).

\bibitem{YellowBook} P. Di Francesco, P. Mathieu, D. Senechal, {\it Conformal field theory}, Springer, New York, 1997.


\bibitem{GST} A.M.~Gainutdinov, H.~Saleur and I.Yu.~Tipunin, \textit{$W$-algebras in XXZ spin chains at roots of unity}, {\tt arXiv:1212.1378}.


\bibitem{DJS} J.~Dubail, J.L.~Jacobsen and H.~Saleur, {\it Conformal field theory at central charge c=0: A measure of the indecomposability (b) parameters}, Nucl. Phys. B {\bf 834}, 399--422 (2010). 

\bibitem{VJS} R.\ Vasseur, J.L.\ Jacobsen and H.\ Saleur,
  {\em Indecomposability parameters in chiral logarithmic
  conformal field theory},
 Nucl.\ Phys.\ B {\bf 851}, 314--345 (2011).

\bibitem{VGJS} R.\ Vasseur, A.M.\ Gainutdinov, J.L.\ Jacobsen and H.\ Saleur,
  {\em Puzzle of bulk conformal field theories at central charge $c=0$},
  Phys.\ Rev.\ Lett. {\bf 108}, 161602 (2012).
  

\bibitem{Kausch} H.G.~Kausch, \textit{Curiosities at $c=-2$}, {\tt hep-th/9510149};  \textit{Symplectic fermions}, Nucl. Phys. B {\bf 583} 513--541 (2000).

\bibitem{Rohsiepe} F.~Rohsiepe,  {\em On reducible but indecomposable representations of the Virasoro algebra}, {\tt hep-th/9611160}.

\bibitem{KytolaRidout} K.~Kyt\"ol\"a and D.~Ridout, {\it On Staggered Indecomposable Virasoro Modules}, J. Math. Phys. {\bf 50}, 123503 (2009). 

\bibitem{GK1} M. Gaberdiel and H. Kausch, \textit{Indecomposable fusion
  products}, Nucl. Phys. B477 (1996) 293-318.

\bibitem{JR} J.~Rasmussen, \textit{Classification of Kac representations in the 
 logarithmic minimal models $\mathcal{LM}(1,p)$}, Nucl. Phys. B {\bf 853}, 404--435 (2011).


\bibitem{RC-paper} A.\ Rocha-Caridi, {\em Vacuum vector representations of the Virasoro algebra}, in {\em Vertex Operators
in Mathematics and Physics}, Eds.\ J.\ Lepowsky, S.\ Mandelstam \& I.\ Singer, Publ.\ Math.\ Sciences, (Springer-Verlag, New York, 451, 1985).

\bibitem{MathieuRidout1} P.~Mathieu and D.~Ridout, {\it Logarithmic M(2,p) Minimal Models, their Logarithmic Couplings, and Duality}, Nucl. Phys. B {\bf 801}, 268 (2008). 


\bibitem{Gurarie2} V.~Gurarie, {\it c-Theorem for Disordered Systems}, Nucl. Phys. B {\bf 546}, 765 (1999). 

\bibitem{GurarieLudwig2}   V.~Gurarie and A.W.W.~Ludwig, {\it Conformal algebras of two-dimensional disordered systems}, J. Phys. A {\bf 35}, L377 (2002). 

\bibitem{RidoutBulk} D.\ Ridout,  {\em Non-chiral logarithmic couplings for the Virasoro algebra}, 
J.\ Phys.\ A: Math.\ Theor.\ {\bf 45}, 255203 (2012).
  
 

\bibitem{MartinSaleur} P.P. Martin and H. Saleur,  \textit{The blob algebra
and the periodic Temperley-Lieb algebra}, Lett. Math. Phys. {\bf 30}, 189--206 (1994).

\bibitem{JSCombBlob} J.L.~Jacobsen and H.~Saleur, {\it Combinatorial aspects of boundary loop models}, J. Stat. Mech. P01021 (2008). 

\bibitem{JSBlob} J.L.~Jacobsen and H.~Saleur, {\it Conformal boundary loop models}, Nucl. Phys. B {\bf 788}, 137--166 (2008). 

\bibitem{Martin} P.P.\ Martin and D.\ Woodcock,
 {\em On the structure of the blob algebra}, J.\ Algebra {\bf 225}, 957--988 (2000). 

\bibitem{[K-first]} H.G.~Kausch, \textit{Extended conformal algebras
    generated by a multiplet of primary fields}, Phys. Lett. B~259
  (1991) 448.

\bibitem{GSchW} M.B.~Green, J.H.~Schwarz, and E.~Witten, \textit{Superstring Theory, Volume 2: Loop Amplitudes,
Anomalies, and Phenomenology}, Cambridge University, Cambridge, 1987.

\bibitem{SaleurSchomerus} H. Saleur and V. Schomerus, \textit{On the $SU(2|1)$ WZW model and its statistical mechanics applications}, Nucl. Phys. B775 (2007) 312.

\bibitem{MartinSaleur1} P.P. Martin and H. Saleur, \textit{On an
  algebraic approach to higher-dimensional statistical mechanics}, Comm. Math. Phys. 158 (1993) 155.

\bibitem{GL} J.J. Graham and G.I. Lehrer,  \textit{The representation
theory of affine Temperley-Lieb algebras}, L'Ens. Math. 44 (1998) 173.

\bibitem{Cardy01} J. Cardy, {\em The stress tensor in quenched random systems}, in
A. Cappelli and G. Mussardo (eds.), {\em Statistical field theories},
NATO Sci. Ser. II: Math. Phys. Chem. {\bf 73}, 215-222 (2002); {\tt
arXiv:cond-mat/0111031}.

\bibitem{VJS12} R. Vasseur, J.L. Jacobsen and H. Saleur, {\em Logarithmic
observables in critical percolation}, J. Stat. Mech. L07001 (2012); {\tt
arXiv:1206.2312}.

\bibitem{VJS13} R. Vasseur, J.L. Jacobsen and H. Saleur, {\em Watermelon operators
and logarithmic correlations in the critical Potts model in $d$ dimensions}, in
preparation (2013).

\bibitem{Coniglio82} A. Coniglio, {\em Cluster structure near the percolation threshold},
J. Phys. A: Math. Gen. {\bf15}, 3829 (1982).

\bibitem{Polyakov70} A.M. Polyakov, Pisma ZhETP {\bf 12}, 538 (1970)
[JETP Lett. {\bf 12}, 381 (1970)].








\bibitem{[HLZ]} Y.-Z.~Huang, J.~Lepowsky, and L.~Zhang,
  \textit{Logarithmic tensor category theory for generalized modules
    for a conformal vertex algebra, I-VIII},
 {\tt arXiv:1012.4193; arXiv:1012.4196; arXiv:1012.4197; arXiv:1012.4198; arXiv:1012.4199; arXiv:1012.4202,  arXiv:1110.1929,  arXiv:1110.1931.}

\bibitem{Nekrasov} E. Frenkel, A. Losev, N. Nekrasov, {\it Instantons beyond topological theory II}, {\tt arXiv:0803.3302}.

\bibitem{MooreSeiberg} G. Moore and N. Seiberg, Lectures on RCFT, in: Physics, Geometry, and Topology, H.C. Lee,
ed. (Plenum Press, New York 1990), p. 263.

\bibitem{[AM1]} D.~Adamovi\'c, A.~Milas, 
\textit{On W-algebras associated to (2,p) minimal models and their
  representations},  {\tt arXiv:0908.4053}.
\bibitem{[TW]}  A.~Tsuchiya, S.~Wood,
\textit{The tensor structure on the representation category of the $W_p$ triplet algebra}, {\tt arXiv:1201.0419}.
 














\end{thebibliography}
\end{document}